\documentclass[aps,prb,twocolumn,superscriptaddress]{revtex4-2}
\usepackage{epsfig,amsopn,siunitx}
\usepackage{graphicx}
\usepackage{array}
\usepackage[margin=1in]{geometry}
\usepackage{color}
\usepackage{hyperref}
\hypersetup{
    colorlinks=true,
    linkcolor=magenta,
    citecolor=magenta,
    urlcolor=magenta
}
\usepackage{amsmath,amssymb}
\usepackage{enumerate}
\newcommand\bea{\begin{eqnarray}}
\newcommand\eea{\end{eqnarray}}
\newcommand\beq{\begin{equation}}
\newcommand\eeq{\end{equation}}
\newcommand\non{\nonumber}
\newcommand{\noi}{\noindent}

\newcommand{\mcS}{\mathcal{S}}
\newcommand{\mcD}{\mathcal{D}}
\newcommand{\mcO}{\mathcal{O}}

\def\nn{\nonumber}
\def\f{\frac}
\def\al{\alpha}

\def\ep{\epsilon}

\def\ga{\gamma}

\def\lm{\lambda}

\def\ra{\rangle}

\def\inf{\infty}

\begin{document}

\title{Topological zero-reflection points in multi-terminal quantum wire junctions}
\author{ Abhiram Soori} 
\email{abhirams@uohyd.ac.in}
\affiliation{School of Physics, University of Hyderabad, Prof. C. R. Rao Road, Gachibowli, Hyderabad 500046, India}
\author{ Udit Khanna} 
 \affiliation{Theoretical Physics Division, Physical Research Laboratory, Navrangpura, Ahmedabad 380009, India}
 \author{Diptiman Sen}
 \affiliation{Center for High Energy Physics, Indian Institute of Science, Bengaluru 560094, India}
 
\begin{abstract}
We study scattering in noninteracting multi-terminal quantum wire junctions and show that junctions with dihedral symmetry can exhibit exact zero-reflection points for $N \ge 4$ terminals. By analyzing the scattering matrix, we identify these reflectionless points in the $(E,t')$ parameter space, where $E$ is the incident particle energy and $t'$ is the junction hopping amplitude. These points exhibit an even-odd dependence on $N$ and converge asymptotically to a common limiting value in the large-$N$ limit. We show that the reflectionless points are characterized by an integer winding number associated with the phase of the reflection amplitude, providing a topological description for their stability against weak on-site disorder. We also consider junctions with broken time-reversal symmetry and find that a magnetic flux can induce additional reflectionless points, including for the $N = 3$ case. For a four-terminal junction threaded by a $\pi$-flux, we identify a unique parameter regime in which the reflection amplitude vanishes over the entire energy band. Finally, we discuss experimental signatures through the behavior of Friedel oscillations and examine the stability of these reflectionless points in the presence of weak interactions.
\end{abstract}

\maketitle

\section{Introduction}
\label{sec1}

The study of electron transport in quantum wire junctions has long been an important theme in condensed matter physics, spanning both noninteracting and interacting regimes~\cite{rganesh2025,lal2002,chamon2003,oshikawa2006,soori2011epl}. In the noninteracting limit,  a junction of $N$ identical wires meeting at a single point exhibits fundamental constraints on transport; for instance, a three-wire junction possesses a reflection probability bounded from below by $1/9$. 
 In fact, the minimum reflection probability increases with $N$ and asymptotically approaches unity in the large-$N$ limit~\cite{rganesh2025}.

Such constraints prompt the fundamental question of whether it is possible to engineer a junction of $N$ wires 
that supports reflectionless transport. Assuming the wires are identically connected to the junction, for even $N$, one may envision a junction that effectively decouples the junction into pairs of smoothly connected wires. However, for odd $N$, this possibility does not exist. Therefore, a generic reflectionless junction (with any $N$) must involve partial transmission to the other wires.

The search for reflectionless scattering modes is of significant interest across diverse physical platforms. In electromagnetism, such modes have been extensively investigated within the context of waveguides~\cite{stone2020}, while in condensed matter, they characterize the topological phase boundaries of the Andreev bound state spectrum in multi-terminal Josephson junctions~\cite{belzig2025}. In two-terminal setups, reflectionless points can lead to quantized charge pumping when the system parameters are adiabatically driven along a closed contour encircling these points~\cite{avron2001,entin2002,avron2004}. On the experimental front, the successful fabrication of multi-terminal quantum wire architectures~\cite{goni1992,Li1999} provides a viable platform for exploring such transport phenomena in the solid state.

In this work, we demonstrate that non-trivial reflectionless transport can be exactly realized in noninteracting multi-terminal quantum wire junctions. We identify specific coordinates in the parameter space where the reflection amplitude vanishes identically, and establish that these zeros are protected by a topological invariant. This underlying topological structure ensures the robustness of the reflectionless condition against weak perturbations, providing a rigorous framework for designing perfectly transmitting nodes in complex quantum networks. Furthermore, we reveal that a four-terminal junction threaded by a $\pi$-flux exhibits a broadband zero-reflection amplitude, yielding perfect transparency across the entire incident energy spectrum. Finally, we analyze the stability of these phenomena under weak interactions and propose an experimental detection scheme based on real-space Friedel oscillations.

The remainder of this paper is organized as follows. In Sec.~\ref{sec2}, we present general symmetry considerations of the scattering matrix for $N$-terminal junctions, establishing the group-theoretic prerequisites for reflectionless transport. Section~\ref{sec4} details the microscopic tight-binding Hamiltonian and the exact scattering formalism used in our analysis. In Sec.~\ref{sec5}, we provide explicit analytical and numerical results for the zero-reflection coordinates across various $N$-wire geometries, including their asymptotic behavior in the large-$N$ limit. The underlying topological structure of these reflectionless points, characterized by an integer winding number, is established in Sec.~\ref{sec6}. We then examine the robustness of this topological protection against weak on-site disorder in Sec.~\ref{sec7}, followed by a discussion of zero reflection points in junctions with unequal hopping amplitudes in Sec.~\ref{sec:uneqhop}. Then, we explore the generation of reflectionless modes via time-reversal symmetry breaking in Sec.~\ref{sec8}. Building on this, Sec.~\ref{sec:pi4} investigates a four-terminal junction threaded by a $\pi$-flux, revealing the emergence of a broadband zero-reflection amplitude across the entire incident energy spectrum. In Sec.~\ref{sec:N2}, we briefly revisit the minimal $N=2$ system to illustrate that these topological features manifest even in a simple two-terminal geometry with local impurities. 
Section~\ref{sec9} discusses the physical implications of our findings, specifically proposing an experimental detection scheme based on real-space Friedel oscillations and analyzing the stability of zero-reflection fixed points under weak inter-particle interactions. Additionally, we examine the behavior of these transport signatures at finite temperatures. Finally, we summarize our results and conclude in Sec.~\ref{sec10}.

\section{General considerations}
\label{sec2}

\begin{figure}[t]
\centering
\includegraphics[width=\columnwidth]{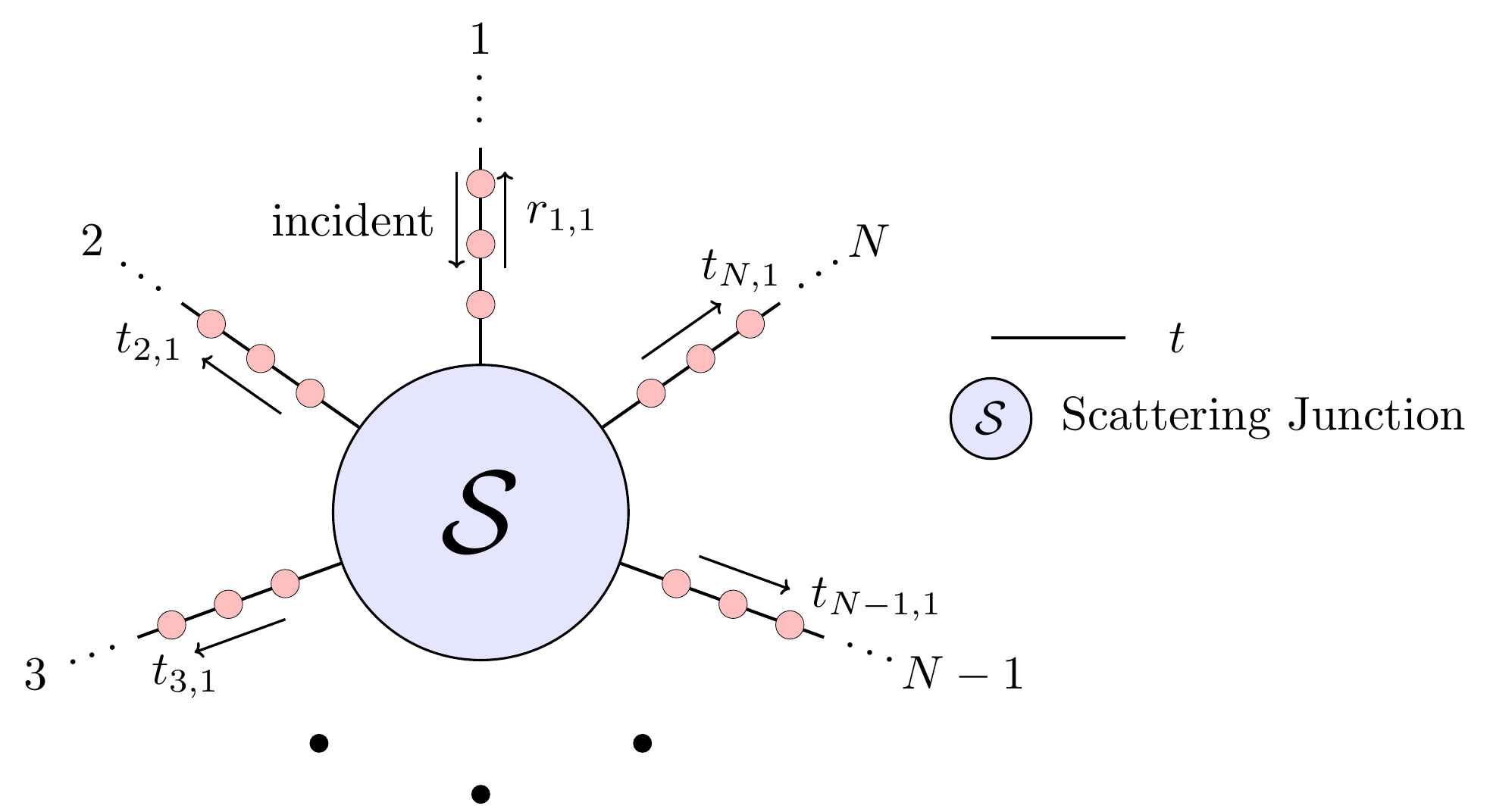}
\caption{Generic scattering junction $\mathcal{S}$ coupled to $N$ tight-binding wire leads.}
\label{fig:generic_junction}
\end{figure}

Ballistic transport through a junction of $N$ quantum wires is governed by the scattering matrix $\mathcal{S}$, that characterizes the junction as a function of the energy of the incoming electrons; in this work we will ignore the
spin of the electrons for simplicity. The conservation of probability ensures that $\mcS$ must be unitary at all energies (we ignore the possibility of inelastic scattering). Then we can examine the existence of reflectionless points in general $N$-terminal junctions very generally by studying $N \times N$ unitary matrices before considering a specific geometry of the junction in detail~\cite{hamermesh}. 

To set the stage for our analysis, we assume that $N$ identical quantum wires, labeled as $n = 1, 2, \dots N$, are connected to the junction with the same tunneling amplitude as shown in Figure~\ref{fig:generic_junction}. Then any of the wires (say, $n = 1$) can be chosen as the incident wire without loss of generality. In this case, the transmission amplitudes depends only on the symmetries of the junction. In this work, we shall focus our attention to time reversal symmetric junctions. Then there are two geometric configurations possible: $\mcS$ may be invariant under all elements of (i) the symmetric group on $N$ wires $S_{N}$, or (ii) the dihedral group corresponding to the $N$-sided polygon $D_{N}$. Breaking time-reversal symmetry, say by introducing a flux through the junction, would reduce the symmetry to just the cyclic group of $N$ elements $C_{N}$. 

\subsection{Full Symmetric Group}

In this case, the junction is fully symmetric and immune to the geometric arrangement of the wires. Then all the transmission amplitudes are invariant under any pair-wise permutation $(i,j)$. Hence, $\mcS$ comprises only two complex numbers $r$ (the reflection amplitude) and $t$ (the transmission amplitude). Then the most generic form of the scattering matrix for such a junction is, 
\begin{align}
\mcS = \left( \begin{array}{ccccc} 
r & t & t & \dots & t \\
t & r & t & \dots & t \\
\vdots & & \ddots & & \vdots \\
t & t & t & \dots & r
\end{array} \right)
\end{align}
Without loss of generality, we can assume that $r$ is real by adding an overall phase to $\mcS$ which does not appear in any physical observables corresponding to a single junction. Imposing unitarity on $\mcS$, we find that $\mcS$ is characterized by a single angle $\theta$. We may write $r, t$ as, $r = \cos \theta, \, t = e^{i \phi} \frac{1}{\sqrt{N-1}} \sin \theta $, which satisfies, 
\begin{align}
\tan \theta = -2\frac{\sqrt{N-1}}{N-2} \cos \phi
\end{align}
A reflectionless point, $r = 0$, would correspond to $\theta = \pm \frac{\pi}{2}$ or $\tan \theta = \pm \infty$. However, $\cos \phi$ is bounded between $-1$ and $1$, and therefore $\tan \theta$ must be finite for all $N > 2$. Therefore, a fully symmetric multiterminal junction does not have any reflectionless points. In fact, the reflection coefficient, $r = \cos \theta = 1 - \frac{2}{N}$, approaches $1$ for $N \gg 1$, implying that the junction becomes almost perfectly reflecting for large $N$. We note that this was demonstrated for the case of $N$ wires connected to a single site in Ref.~\cite{rganesh2025}. Here, we show that the result holds for any fully symmetric junction, irrespective of the microscopic structure.

\subsection{Dihedral Group}

Next, we consider junctions that have a somewhat lower symmetry -- instead of being invariant under any permutation of the incoming leads, we assume that the junction is invariant under the elements of the dihedral group $D_{N}$ which describes the symmetries of an $N$-sided polygon. The generators of the group are rotation by $\frac{2\pi}{N}$ and a reflection. In the context of a multiterminal junction in which the leads are ordered as $1,2,3,\dots,N-1,N$, the former corresponds to a cyclic shift of the labels $(1,2,\dots,N) \rightarrow (N,1,2,\dots,N-1)$, while the reflection about one of the leads (say about lead 1) corresponds to an interchange of the form $i \Leftrightarrow N+2-i$ (for $i > 1$), i.e., $(1,2,\dots,N-1,N) \rightarrow (1,N,N-1,\dots,3,2)$. In this case, we have to analyze each $N$ separately. 

\subsubsection{N = 3}

Since the dihedral group $D_{3}$ is isomorphic to the symmetric group $S_{3}$, we note that our results for the fully symmetric case immediately carry over (for a junction with three leads) if time-reversal symmetry is not broken. Therefore, such a junction does not have any reflectionless points. In fact, the minimal reflection amplitude is $r = 1 - \frac{2}{3} = \frac{1}{3}$. The case of time-reversal symmetry broken junctions is considered in section~\ref{sec8}.

\subsubsection{N = 4}
\label{N4-1}

\begin{align}
\mcS = \left( \begin{array}{cccc} 
r & t_{1} & t_{2} & t_{1} \\
t_{1} & r & t_{1} & t_{2} \\
t_{2} & t_{1} & r & t_{1} \\
t_{1} & t_{2} & t_{1} & r 
\end{array} \right)
\end{align}
As before, we assume (without loss of generality) that $r$ is real, implying that there are 5 real parameters in $\mcS$. Imposing unitarity on $\mcS$, gives 3 equations of the form: $\mcD = 1, \mcO_{1} = 0, \mcO_{2}= 0$, where, 
\begin{align}
\mcD &= r^{2} + |t_{2}|^{2} + 2|t_{1}|^{2}, \label{eq:D4} \\
\mcO_{1} &= 2 \,{\rm Re} \big[r t_{1}  +  t_{1} t_{2}^{*} \big], \label{eq:O4a} \\
\mcO_{2} &= 2 \,{\rm Re} \big[r t_{2} \big] + 2\,|t_{1}|^{2}, \label{eq:O4b} 
\end{align}
These equations can be satisfied by parameterizing the amplitudes as $r = \cos \theta$, $t_{2} = e^{i \phi_{2}} \sin \theta \cos \psi$, and $t_{1} = e^{i \phi_{1}} \sin \theta \sin \psi/\sqrt{2}$. Substituting these expressions into Eqs.~\eqref{eq:O4a} and \eqref{eq:O4b} yields two relations among the four angular parameters. Imposing the zero-reflection condition, $r=0$, requires $\sin\psi=0$ and $\cos(\phi_1-\phi_2)=0$, which consequently yields $t_1=0$ and $t_2=e^{i\phi_2}$. Physically, this indicates that wires $1$ and $3$ are perfectly coupled with vanishing reflection, while scattering from wire $1$ into wires $2$ and $4$ is completely suppressed.

\subsubsection{N = 5}

Consider a 5-terminal junction with the dihedral symmetry $D_{5}$. We label the wires $n = 1, 2, \dots, 5$ and assume that $n = 1$ is the incident wire. Due to the symmetry, there are 2 independent transmission amplitudes: $t_{2,1} = t_{5,1} = t_{1}$ and $t_{3,1} = t_{4,1} = t_{2}$. Then the S-matrix takes the form, 
\begin{align}
\mcS = \left( \begin{array}{ccccc} 
r & t_{1} & t_{2} & t_{2} & t_{1} \\
t_{1} & r & t_{1} & t_{2} & t_{2} \\
t_{2} & t_{1} & r & t_{1} & t_{2} \\
t_{2} & t_{2} & t_{1} & r & t_{1} \\
t_{1} & t_{2} & t_{2} & t_{1} & r
\end{array} \right)
\end{align}
As before, we assume (without loss of generality) that $r$ is real, implying that there are 5 real parameters in $\mcS$. Imposing unitarity on $\mcS$, gives 3 equations of the form: $\mcD = 1, \mcO_{1} = 0, \mcO_{2}= 0$, where, 
\begin{align}
\mcD &= r^{2} + 2|t_{2}|^{2} + 2|t_{1}|^{2}, \label{eq:D5} \\
\mcO_{1} &= 2 {\rm Re} \big(r t_{1} \big) + 2 {\rm Re} \big(t_{1} t_{2}^{*} \big) + |t_{2}|^{2}, \label{eq:O5a} \\
\mcO_{2} &= 2 {\rm Re} \big(r t_{2} \big) + 2 {\rm Re} \big(t_{1} t_{2}^{*} \big) + |t_{1}|^{2}, \label{eq:O5b} 
\end{align}
Equation (\ref{eq:D5}) can be satisfied if we parameterize the amplitudes as, $r = \cos \theta, \, t_{2} = e^{i \phi_{2}} \frac{1}{\sqrt{2}} \sin \theta \cos \psi, \, t_{1} = e^{i \phi_{1}} \frac{1}{\sqrt{2}} \sin \theta \sin \psi$. Plugging this in Eqs.~\eqref{eq:O5a} and \eqref{eq:O5b}, we find 2 relations between the 4 angles. Thus the scattering matrix can be parameterized by 2 real parameters. 

A reflectionless point, $r = 0$, corresponds to $\theta = \pm \frac{\pi}{2}$ or $\tan \theta = \pm \infty$. Unlike the case of $N = 3$, here we find that it is possible to satisfy unitarity and $r = 0$ simultaneously. For the reflectionless point, we find, 
\begin{align}
r &= 0, \, t_{2} = \frac{e^{i \phi_{2}}}{2} \, \text{ and } \, t_{1} = e^{i 2\pi/3} t_{2}. 
\end{align}
To put this solution in perspective, we note that the solution above proves that a reflectionless point exists in the space of \textit{all} allowed S-matrices with $D_{5}$ symmetry. 
Whether this is realized in a given microscopic realization for any set of parameters would depend on details of the specific junction. 

\subsubsection{N = 6}

Consider a 6-terminal junction with the dihedral symmetry $D_{6}$. As before we label the wires $n = 1, 2, \dots, 6$ and assume that $n = 1$ is the incident wire. Due to the symmetry, there are 3 independent transmission amplitudes: $t_{2,1} = t_{6,1} = t_{1}$, $t_{3,1} = t_{5,1} = t_{2}$ and $t_{4,1} = t_{3}$. Then the S-matrix takes the form, 
\begin{align}
\mcS = \left( \begin{array}{cccccc} 
r & t_{1} & t_{2} & t_{3} & t_{2} & t_{1} \\
t_{1} & r & t_{1} & t_{2} & t_{3} & t_{2} \\
t_{2} & t_{1} & r & t_{1} & t_{2} & t_{3} \\
t_{3} & t_{2} & t_{1} & r & t_{1} & t_{2} \\
t_{2} & t_{3} & t_{2} & t_{1} & r & t_{1} \\
t_{1} & t_{2} & t_{3} & t_{2} & t_{1} & r
\end{array} \right)
\end{align}
As before, we assume (without loss of generality) that $r$ is real, implying that there are 7 real parameters in $\mcS$. Imposing unitarity on $\mcS$, gives 4 equations of the form: $\mcD = 1, \mcO_{1} = 0, \mcO_{2}= 0, \mcO_{3} = 0$, where, 
\begin{align}
\mcD &= r^{2} + |t_{3}|^{2} + 2|t_{2}|^{2} + 2|t_{1}|^{2}, \label{eq:D6} \\
\mcO_{1} &= 2 {\rm Re} \big(r t_{1} \big) + 2 {\rm Re} \big(t_{2} t_{3}^{*} \big) + 2 {\rm Re} \big(t_{1} t_{2}^{*} \big), \label{eq:O6a} \\
\mcO_{2} &= 2 {\rm Re} \big(r t_{2} \big) + 2 {\rm Re} \big(t_{1} t_{3}^{*} \big) + |t_{1}|^{2} + |t_{2}|^{2}, \label{eq:O6b} \\
\mcO_{3} &= 2 {\rm Re} \big(r t_{3} \big) + 4 {\rm Re} \big(t_{1} t_{2}^{*} \big). \label{eq:O6c} 
\end{align}
Equation (\ref{eq:D6}) can be satisfied if we parameterize the amplitudes as, $r = \cos \theta, \, t_{3} = e^{i \phi_{3}} \sin \theta \cos \psi, \, t_{2} = e^{i \phi_{2}} \frac{1}{\sqrt{2}} \sin \theta \sin \psi \cos \beta, \, t_{1} = e^{i \phi_{1}} \frac{1}{\sqrt{2}} \sin \theta \sin \psi \sin \beta$. Plugging this in Eqs.~\eqref{eq:O6a}
and \eqref{eq:O6c}, we find 3 relations between the 6 angles. Thus the scattering matrix can be parameterized by 3 real parameters. 

A reflectionless point, $r = 0$, corresponds to $\theta = \pm \frac{\pi}{2}$. This may occur in two situations. First, an isolated reflectionless point occurs for $\sin \psi = 0$, which means that $t_1 = t_2 = 0$ and 
$|t_{3}| = 1$. This corresponds to a \textit{trivial} reflectionless point, where the diagonally opposite wire is perfectly transmitting. Such a solution exists for all junctions with even $N$, including for $N = 4$ as we had seen before. 

However, the $N = 6$ case presents an additional \textit{non-trivial family} of reflectionless points, with partial transmission in all wires. This solution for reflectionless points is parameterized by 2 angles,
\bea
t_1 &=& \frac{e^{i\phi}(1-\cos\beta)}{\sqrt{5-4\cos\beta}}, \\
t_2 &=& \frac{ie^{i\phi}\sin\beta}{\sqrt{5-4\cos\beta}}, \\
t_3 &=& -\frac{e^{i\phi}}{\sqrt{5-4\cos\beta}}.
\eea
Note that a special case of this, given by $\beta = 0$, reproduces the reflectionless point 
discussed in the previous paragraph.

While these equations suggest that there is a continuous family of reflectionless points, we emphasize that this only exists in the space of all S-matrices allowed by the symmetry. A given microscopic realization of a 6-terminal junction would only admit a subset of the complete space of S-matrices, and the trivial and/or non-trivial reflectionless points would exist only if the subset intersects this family. 


 \begin{figure}
\includegraphics[width=7cm]{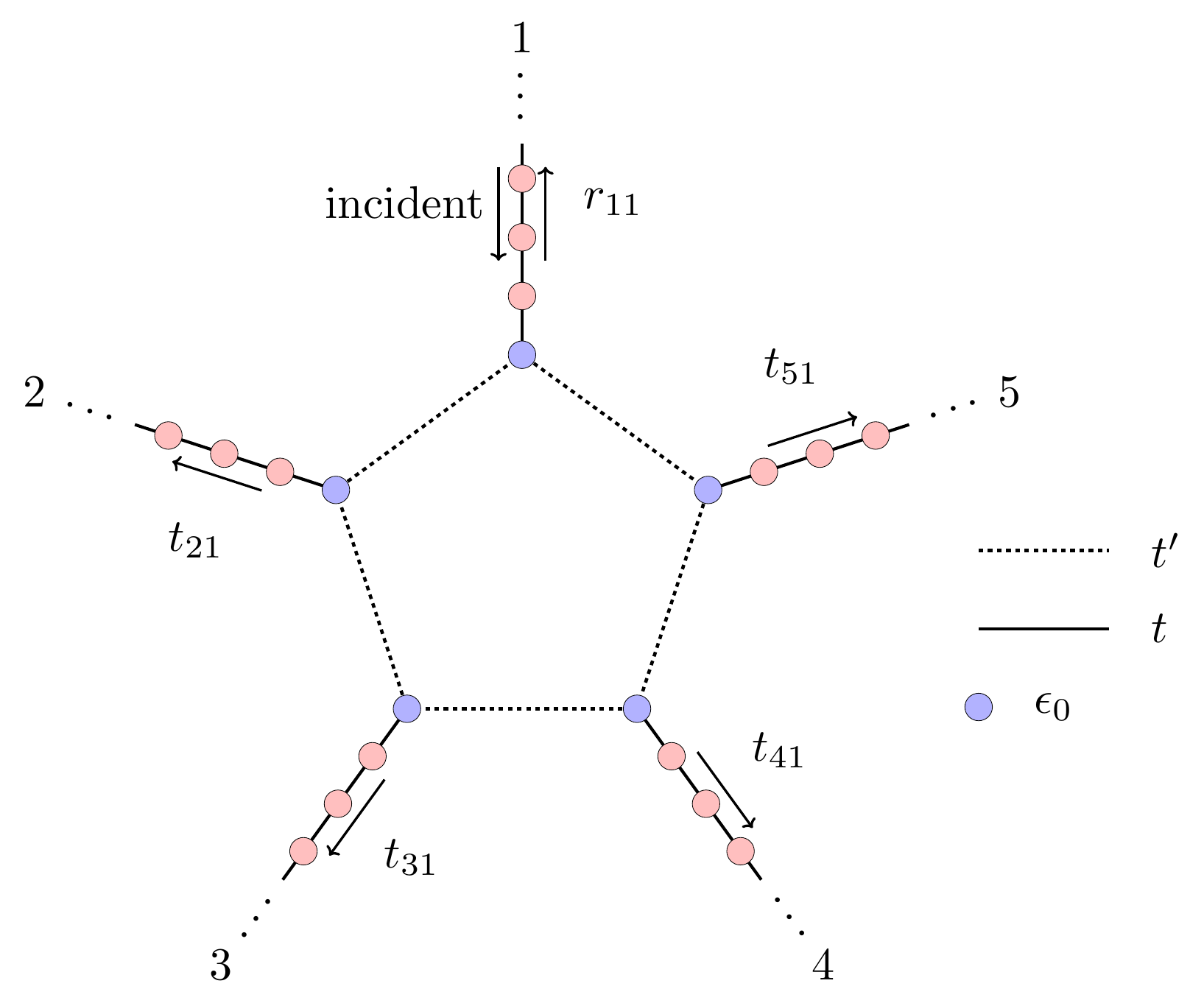}
\caption{Schematic picture of $N$-wire junction with $N=5$ wherein each quantum wire is connected to two other wires. The first site of the chain has an on-site energy $\ep_0$, $t'$ is the hopping amplitude that connects each quantum wire to two other wires and $t$ is the hopping amplitude within each wire.} \label{fig1}
\end{figure}

\section{Polygonal Junctions}
\label{sec4}

 Following the general considerations in the previous section, we consider a specific microscopic model of junctions with dihedral symmetry. Specifically, we consider polygonal junctions (cf. Fig.~\ref{fig1}) in which the junction is composed of $N$-sites arranged as a regular $N$-sided polygon.
The system consists of $N$ semi-infinite quantum wires connected to one another making a junction. The indices for lattice sites in each wire go from $l=0$ to $l=\infty$. The Hamiltonian describing the system is given by 
\bea
H &=& -t ~\sum_{n=1}^{N}\sum_{l=0}^{\infty} ~(c_{n,l+1}^{\dag}c_{c_n,l} + {\rm H.c.})\nn \\ 
	&& -t' ~\sum_{n=1}^{n=N}(c_{n+1,0}^{\dag}c_{n,0}+{\rm H.c.}) \nn\\ &&+ ~\ep_0~ \sum_{n=1}^N c^{\dagger}_{n,0}c_{n,0}~, \label{eq:ham}
\eea
where $c_{n,l}$ corresponds to annihilation of a particle at site $l$ of $n$-th wire, $t$ is the nearest neighbor hopping amplitude in each of the semi-infinite quantum wires, $t'$ is the hopping amplitude that connects site-$0$ of $n$-th wire to site $0$ of $(n+1)$-th wire (note that $N+1\equiv 1$) and $\ep_0$ is the on-site energy for electrons on site $0$ of each of the quantum wires. For simplicity, we set $t=1$. (We will also set $\hbar$
to unity in this paper). For wave functions of
the form $e^{\pm i k l}$ in
the wires, the dispersion is given by 
\beq E(k) ~=~ -~2\cos (ka), \label{disp} \eeq
where $a$ is the lattice spacing (we will henceforth
set $a=1$). We study scattering of a plane wave incident on the junction with energy $E$ from wire $1$. The scattering eigenfunction for such a state has the form:
\bea 
|\psi\ra &=& \sum_{n=1}^{N}\sum_{l=0}^{\infty} ~\psi(n,l)|n,l\ra, ~~{\rm where}~~\nn \\ 
\psi(n,l) &=& \begin{cases}
e^{-ikl} +r_{1,1}e^{ikl} ~~{\rm for}~~n=1, \\ 
t_{n,1}e^{ikl} ~~{\rm otherwise}.
\end{cases}\label{eq:psi}
\eea
Here, $k=\cos^{-1}(-E/2)$ lies in the range 
$0 < k < \pi$. (We do not include the values $k=0$ and $\pi$
since the group velocity $dE/dk = 0$ at those points).
The time-independent Schr\"odinger equation $H|\psi\ra=E|\psi\ra$ for $\psi (n,0)$, with $n=1,2,\cdots,N$ (where $N \ge 4$),
give the equations
\bea && - ~ ~(e^{-ik} + r_{1,1} e^{ik}) - t' ~(t_{2,1} +
t_{N,1}) \non \\
&& + ~\ep_0 ~(1 + r_{1,1}) ~=~ E ~(1 + r_{1,1}), \non \\
&& - ~ ~t_{2,1} e^{ik} ~-~t' ~(1 + r_{1,1}) ~-~ t' ~t_{3,1} 
\non \\
&& +~ \ep_0 ~t_{2,1} ~=~ E ~t_{2,1}, \non \\
&& -~ ~t_{N,1} e^{ik} ~- ~t' ~(1 + r_{1,1}) ~-~ t' t_{N-1,1} 
\non \\
&& +~ \ep_0 ~t_{N,1} ~=~ E ~t_{N,1}, \non \\
&& -~ ~t_{n,1} e^{ik} ~- ~t' ~t_{n-1,1} ~-~ t' ~t_{n+1,1} 
\non \\
&& +~ \ep_0 ~t_{n,1} ~=~ E ~t_{n,1} ~~{\rm for}~~ 3 \le n 
\le N-1. \non \\
\label{cond1} \eea
Using these equations, the scattering coefficients $r_{1,1}$ and $t_{n,1}$ can be calculated. 

We now study the condition for there to be no reflection,
$r_{1,1} = 0$. We find that $r_{1,1} = 0$ will
hold if the following conditions are satisfied:
\bea && -~ t' ~(t_{2,1} + t_{N,1}) ~+~ e^{ik} ~+~ \ep_0 ~=~ 0, \non \\
&& -~t' ~-~ t' ~t_{3,1} ~+~ (e^{-ik} ~+~ \ep_0) ~t_{2,1} ~=~ 0, \non \\
&& - ~t' ~-~ t' ~t_{N-1,1} ~+~ (e^{-ik} ~+~ \ep_0) ~t_{N,1} ~=~ 0, \non \\
&& - ~t' ~t_{n-1,1} ~-~ t' ~t_{n+1,1} ~+~ (e^{-ik} ~+~ \ep_0) ~t_{n,1} \non \\
&& =~ 0 ~~{\rm for}~~ 3 \le n \le N-1. 
\label{cond2} \eea
The last equation in Eq.~\eqref{cond2} has the general solution
\bea t_{n,1} &=& a_+ (\lm_+)^n ~+~ a_- (\lm_-)^n, \non \\
\lm_{\pm} &=& \ga ~\pm~ i ~
\sqrt{1 - \ga^2}~, \non \\
&& {\rm where~} \ga= \frac{e^{-ik} + \ep_0}{2t'}
\label{lpm} \eea
and $a_\pm$ are some constants. We note that 
$\lm_\pm$ satisfy the relation $\lm_+ \lm_- = 1$.

Since Eqs.~\eqref{cond2} are symmetric under the $n \leftrightarrow N+2-n$, we can assume that the solutions of those equations have
the same symmetry, namely,
\beq t_{n,1} ~=~ t_{N+2-n,1} ~~{\rm for}~~ 1 \le n \le N+1, 
\label{sym} \eeq
where we define $t_{1,1} = t_{N+1,1} \equiv \psi (1,0) = 1$.
We now separately discuss the cases where $N$ is even and
odd respectively. 

\noi (i) For $N=2m$, we take
\beq t_{n,1} ~=~ \frac{(\lm_+)^{n-m-1} ~+~ (\lm_+)^{m+1-n}}{
(\lm_+)^{-m} ~+~ (\lm_+)^m}, \label{neven} \eeq
where $1 \le n \le N$, and we have used the relation 
$\lm_- = 1/\lm_+$. We find that Eq.~\eqref{neven} satisfies
the symmetry in Eq.~\eqref{sym} and all the equations in Eq.~\eqref{cond2} provided that
\beq ~\frac{(\lm_+)^{1-m} ~+~ (\lm_+)^{m-1}}{(\lm_+)^{-m}
~+~ (\lm_+)^{m}} ~=~ \ga^*. \label{cond3} \eeq

\noi (ii) For $N=2m+1$, we take
\beq t_{n,1} ~=~ \frac{(\lm_+)^{n-m-3/2} ~+~ (\lm_+)^{m+3/2-n}}{
(\lm_+)^{-m-1/2} ~+~ (\lm_+)^{m+1/2}}, \label{nodd} \eeq
where $1 \le n \le N$, and we have used the relation 
$\lm_- = 1/\lm_+$. We find that Eq.~\eqref{nodd} satisfies
the symmetry in Eq.~\eqref{sym} and all the equations in Eq.~\eqref{cond2} provided that
\beq ~\frac{(\lm_+)^{1/2-m} ~+~ (\lm_+)^{m-1/2}}{(\lm_+)^{-m-1/2}
~+~ (\lm_+)^{m+1/2}} ~=~ \ga^*. \label{cond4} \eeq
Solving Eqs.~\eqref{cond3} and \eqref{cond4}, along with the
expression for $\lm_+$ given in Eq.~\eqref{lpm} gives the
reflectionless points as a function of $N, ~t', ~\ep_0$ and $k$.

In the following sections, we explicitly discuss a number 
of cases where there is no reflection.

\section{Results}
\label{sec5}

In this section, we present our results for
various values of $N, ~t'$ and $k$. We will set $t=1$.

\subsection{$N=4$}
\label{N4-2}

For $N=4$ and $\ep_=0$, the scattering amplitudes take the forms,
\bea 
r_{1,1} &=& -\f{e^{-i2k} ~(1-4t'^2e^{-ik}\cos k)}{1 ~-~ 4e^{-2ik}t'^2}, \nn \\ 
t_{2,1}=t_{4,1} &=& \f{2i t' \sin k}{e^{2ik} ~-~ 4t'^2}, \nn \\
t_{3,1} &=& \f{4it'^2e^{-ik}\sin{k}}{e^{2ik} ~-~ 4t'^2}. 
\eea
It is clear that in the limit $t' \to \inf$ and with $k =\pi/2$, we have $r_{1,1}=t_{2,1}=t_{4,1}\to 0$ and $t_{3,1}\to -1$. 

For general $\ep_0$, the conditions for zero-reflection are 
\bea 
E=2\ep_0, && t'\to \pm \infty \label{eq:N4ZR}
 \eea
 At the zero-reflection points, $t_{3,1}=e^{2ik}$ and $t_{2,1}=t_{4,1}=0$, where $k=\cos^{-1}(-E/2)$.

\begin{figure*}
\includegraphics[width=6.5cm]{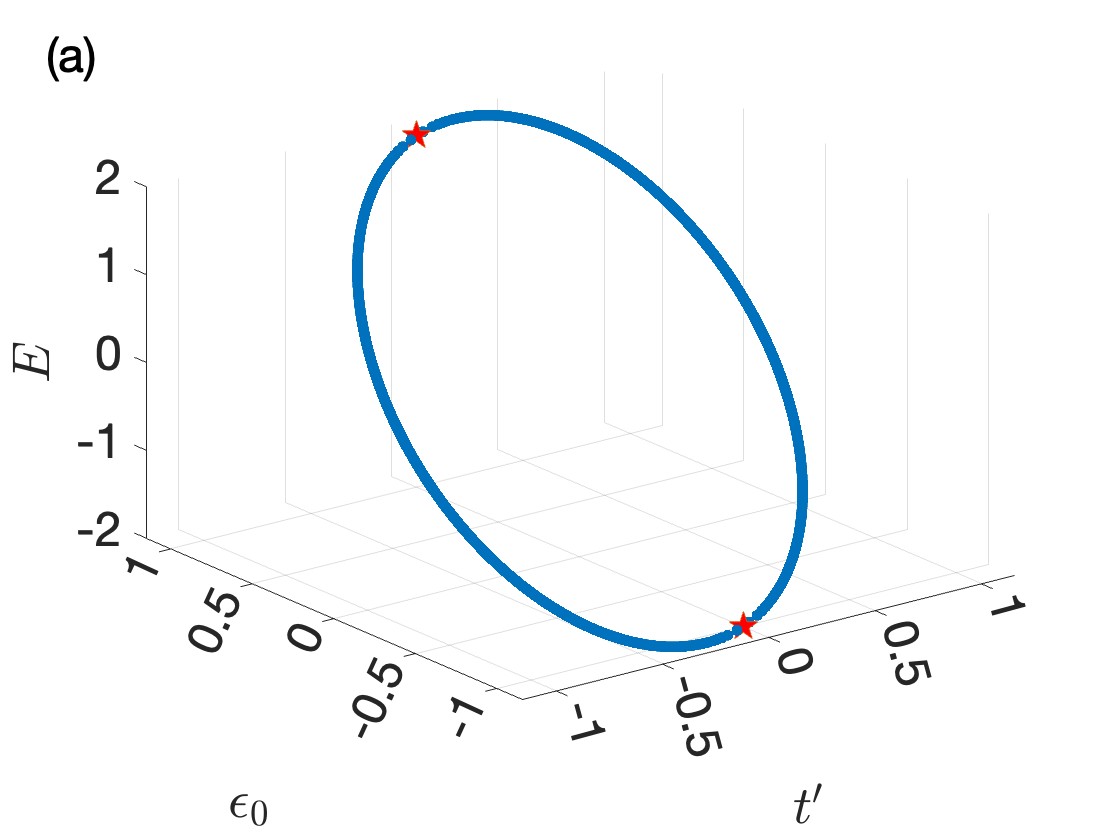}
\includegraphics[width=6.5cm]{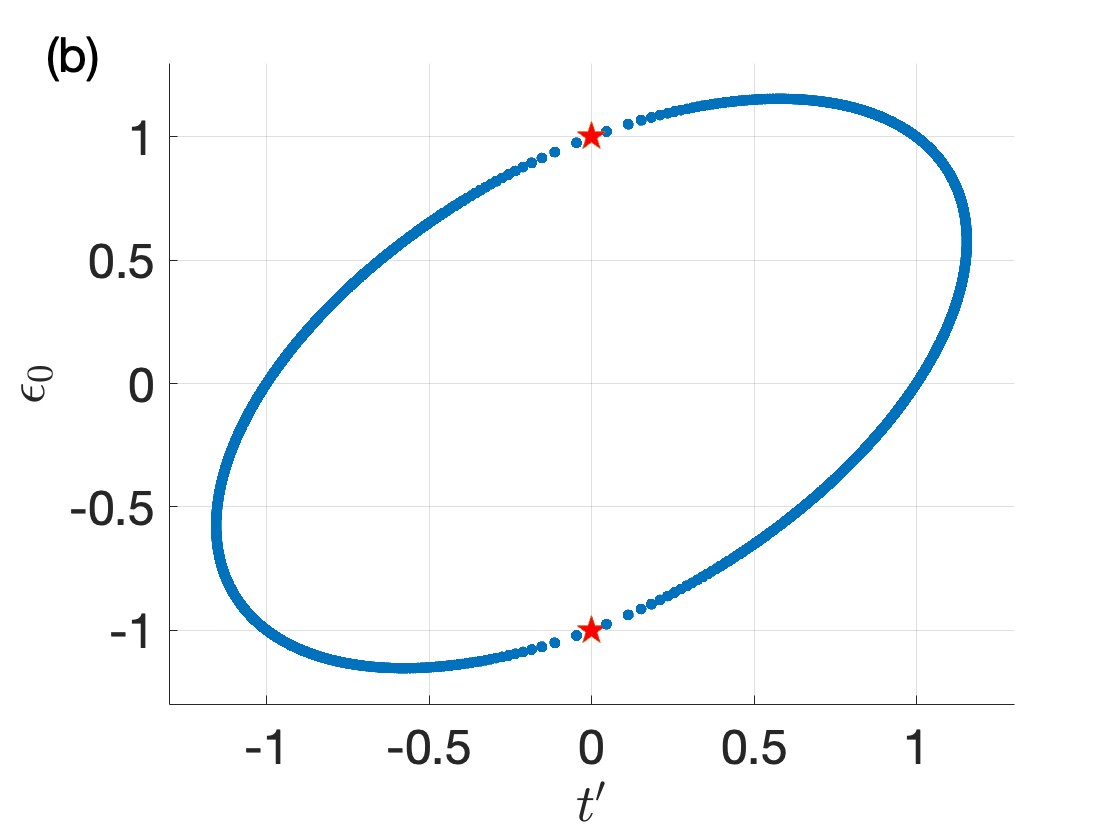}
\includegraphics[width=6.5cm]{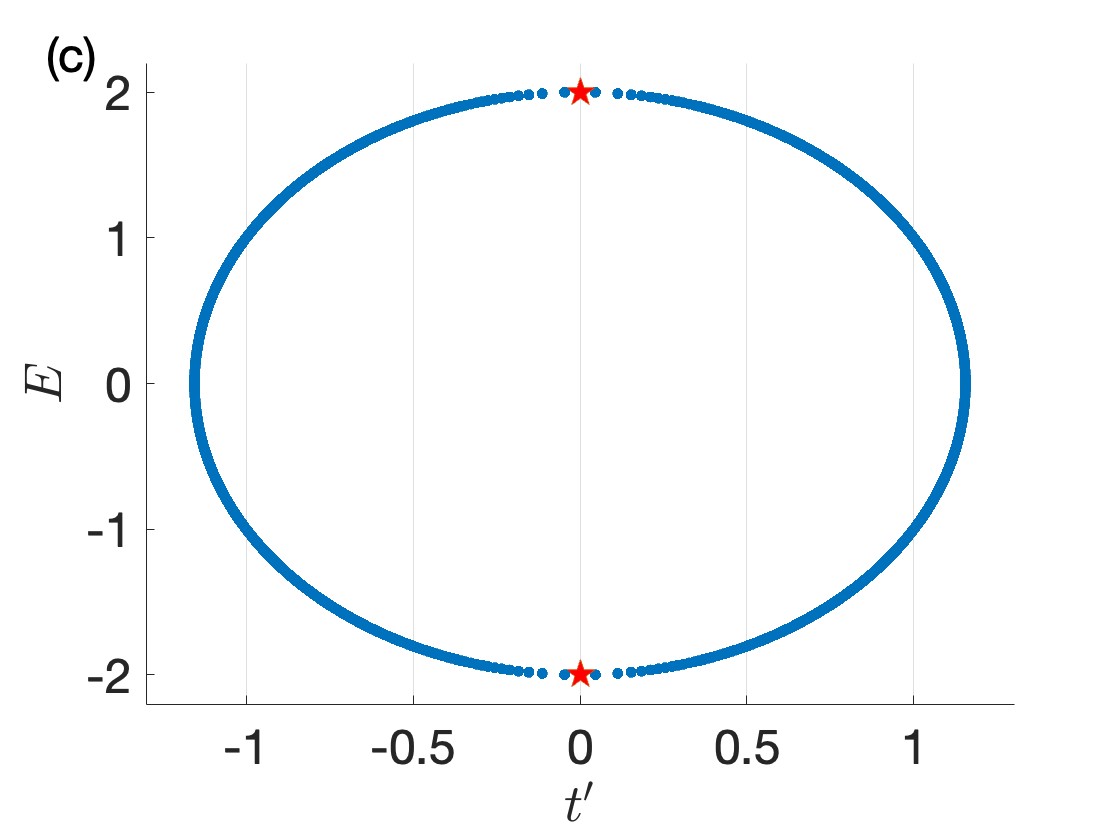}
\includegraphics[width=6.5cm]{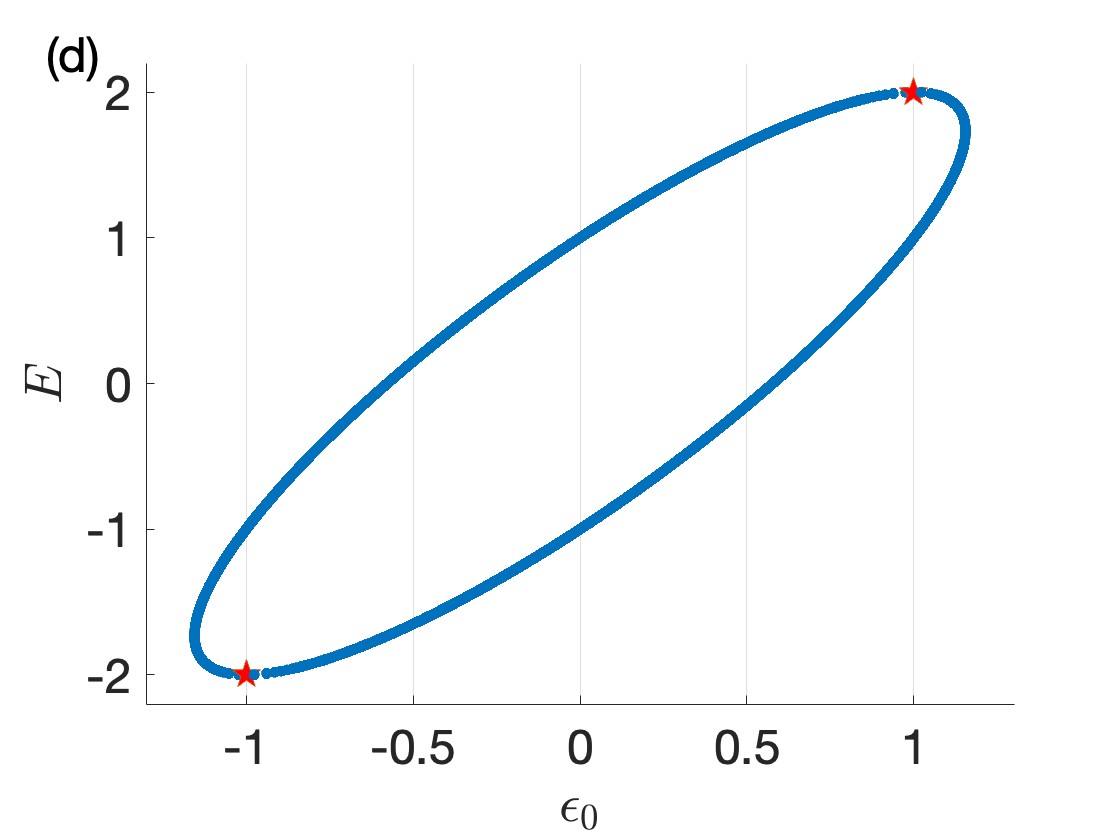}
\caption{Zero-reflection points for a junction with $N=5$ wires. (a) Trajectories of the zero-reflection points in the $(E, t', \epsilon_0)$ parameter space. (b--d) Projections of the zero-reflection points onto the (b) $(t', \epsilon_0)$, (c) $(t', E)$, and (d) $(\epsilon_0, E)$ planes.}
\label{N5}
\end{figure*}

\begin{figure*}
\includegraphics[width=6.5cm]{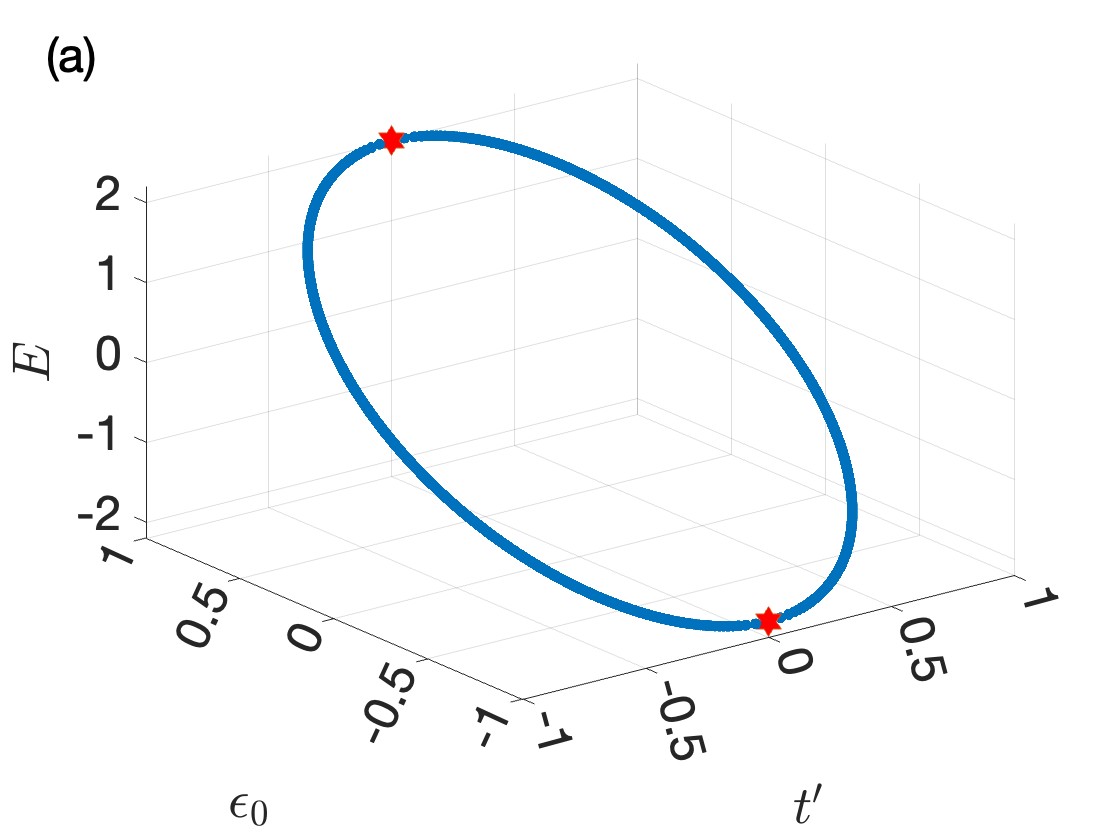}
\includegraphics[width=6.5cm]{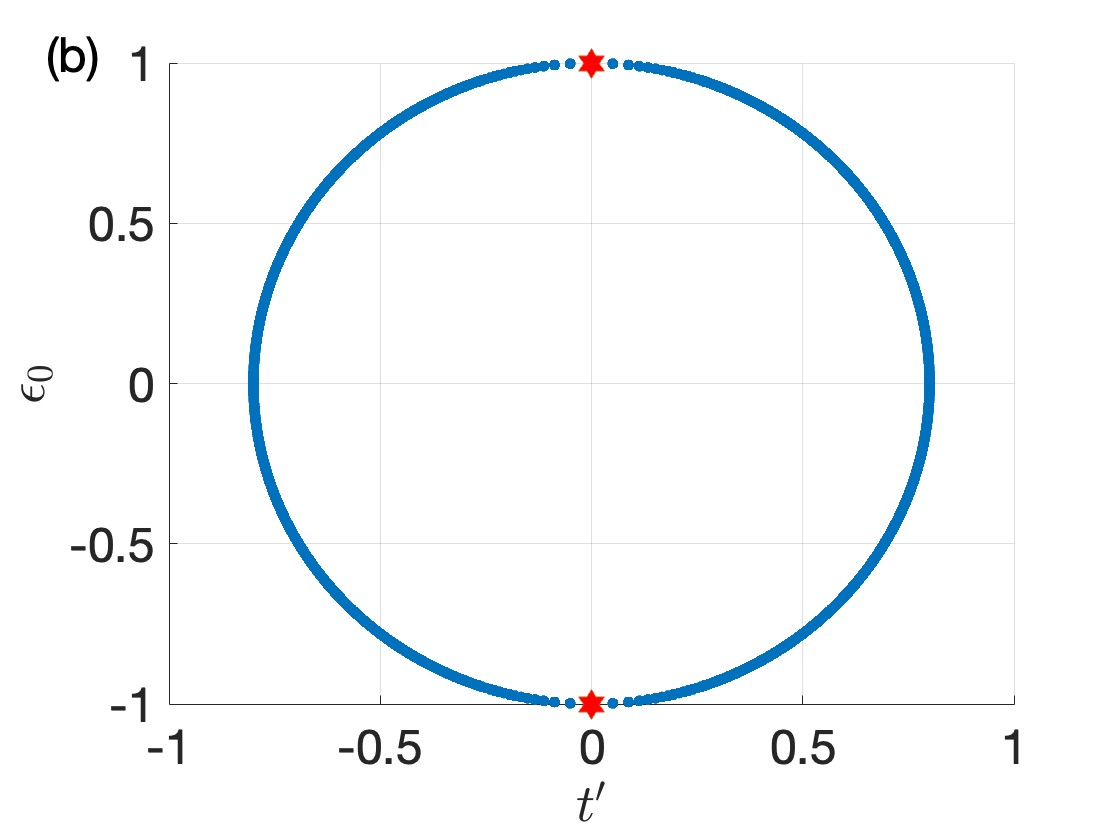}
\includegraphics[width=6.5cm]{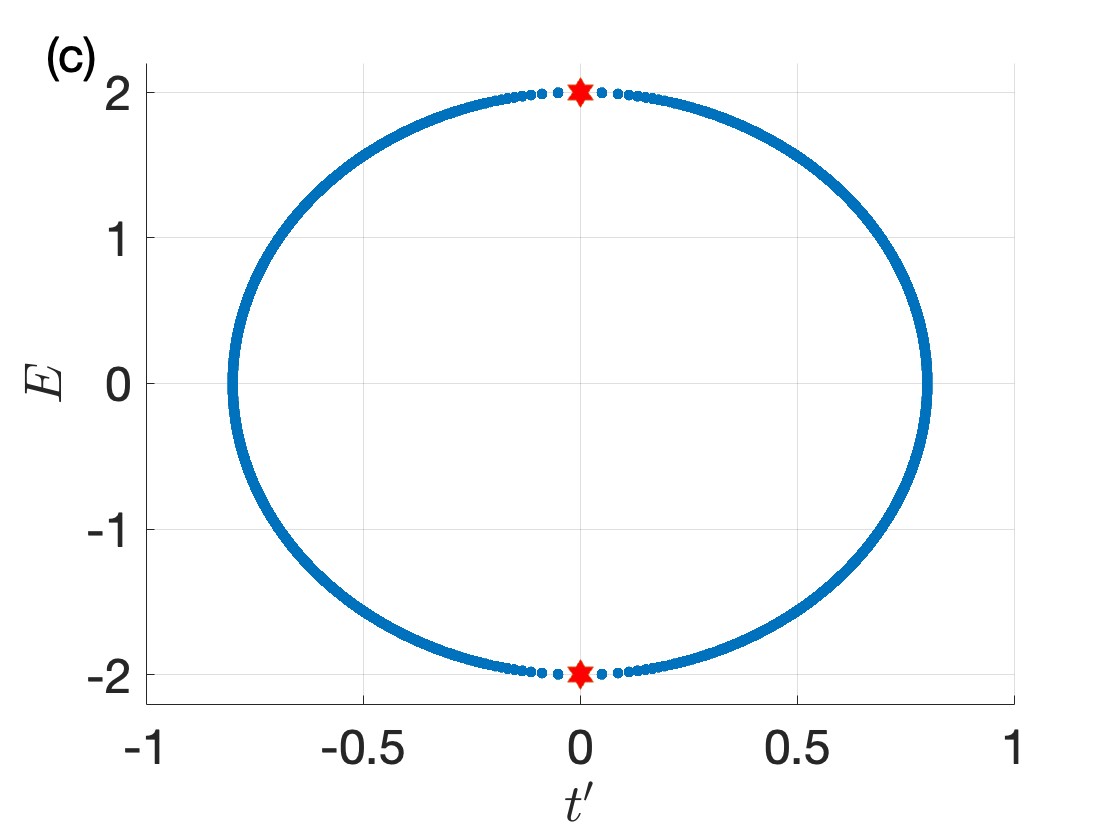}
\includegraphics[width=6.5cm]{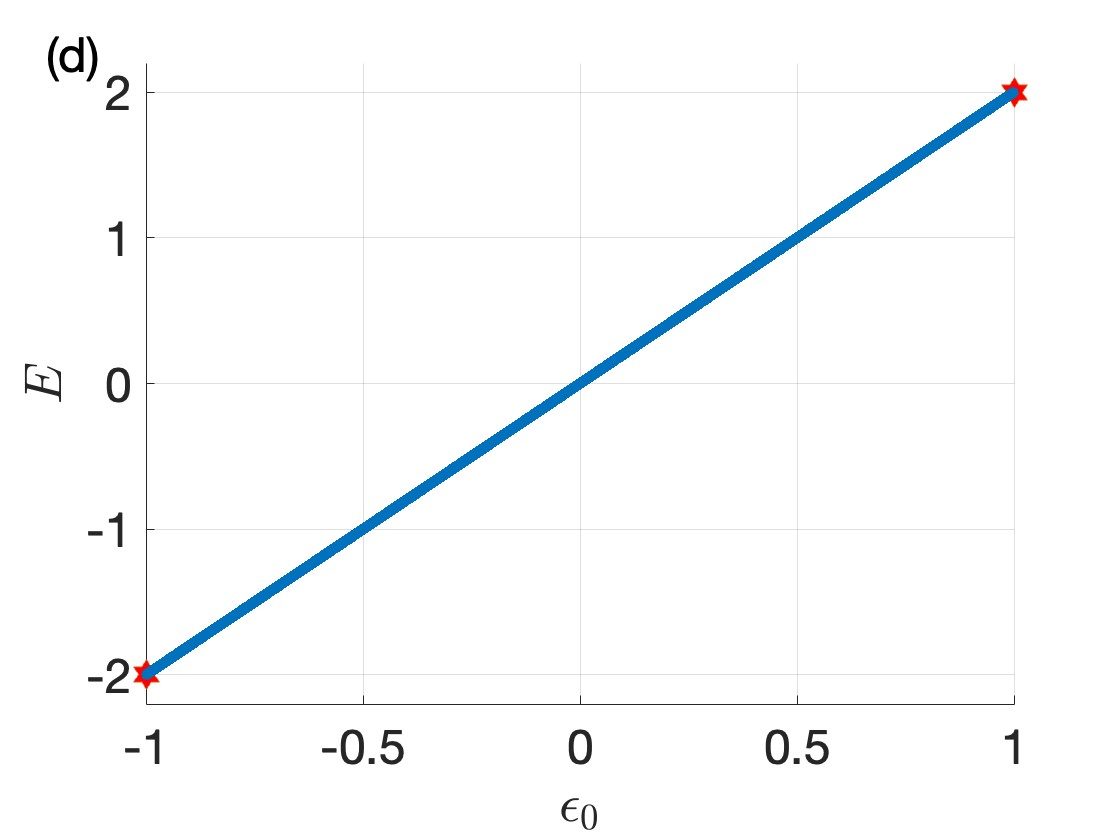}
\caption{ Zero-reflection points for a junction with $N=6$ wires. (a) Trajectories of the zero-reflection points in the $(E, t', \epsilon_0)$ parameter space. (b--d) Projections of these points onto the (b) $(t', \epsilon_0)$, (c) $(t', E)$, and (d) $(\epsilon_0, E)$ planes. Red stars indicate the annihilation points where the zero-reflection condition is lifted and the reflection probability is $1$.} \label{N6}
\end{figure*}

\subsection{$N=5$}

For $N=5$ and $\epsilon_0=0$, we use Eq.~\eqref{cond4} with $m=2$. We
find that the reflection amplitude $r_{1,1}$ is zero 
when either (i) $t'=1$ and $k = \pi/3$, or (ii) 
$t'=-1$ and $k=2 \pi/3$.

For $N=5$ and $\ep_0 \ne 0$, we find from Eq.~\eqref{lpm} and
Eq.~\eqref{cond4} with $m=2$ that there are reflectionless points
when the following conditions are satisfied,
\bea \f{E}{2} &=& \pm ~\sqrt{1 ~-~ \f{3}{4} ~t'^2}, \nn \\
\ep_0 &=& \f{t'}{2} ~\pm~ \sqrt{1 ~-~ \f{3}{4} ~t'^2}, 
\label{ellipse} \eea
and in the second equation, we take the $+ ~(-)$ sign in
front of the square root if $t' < 0 ~(>0)$ respectively.
For $\ep_0 = 0$, we recover the earlier results
for reflectionless points at 
$(t'=1,k=\pi/3)$ and $(t'=-1,k=2\pi/3)$. Eq.~\eqref{ellipse}
implies that the reflectionless points lie on ellipses in terms
of either $(t',\ep_0)$ or $(t',E)$ as shown in 
Fig.~\ref{N5},
\bea 
 \f{t'+E}{2} &=& \ep_0\nn \\ 
\frac{3}{4} ~t'^2 ~+~ \f{E^2}{4} &=& 1. \eea

\subsection{$N=6$}

For $N=6$ and $\epsilon_0=0$, we use Eq.~\eqref{cond3} with $m=3$.
For $k = \pi/2$, we find that the reflection amplitude is zero if $t'$ satisfies the equation
\beq 4 (t')^4 ~-~ (t')^2 ~-~ 1 ~=~ 0. \eeq
The solution of this is given by
\beq t' ~=~ \pm ~\sqrt{\f{1+\sqrt{17}}{8}} ~\simeq~ \pm ~0.8002. \eeq

For $\ep_0\neq 0$, the scattering eigenfunction corresponding to a particle incident from wire-$1$ has the following symmetry: 
$\psi(n,l) =\psi(6-n+2,l),$ which enforces $t_{n,1}=t_{6-n+2,1}$. 
The equations satisfied by the wavefunction $\psi(n,l)$ in Eq.~\eqref{eq:psi} are 
\bea 
E \psi_{1,0} &=& \ep_0\psi_{1,0}-\psi_{1,1}-t'(\psi_{2,0}+\psi_{6,0}) \nn \\ 
E\psi_{2,0} &=& \ep_0\psi_{2,0}-\psi_{2,1}-t'(\psi_{3,0}+\psi_{1,0}) \nn \\ 
E\psi_{3,0} &=& \ep_0\psi_{3,0}-\psi_{3,1}-t'(\psi_{4,0}+\psi_{2,0}) \nn \\ 
E\psi_{4,0} &=& \ep_0\psi_{4,0}-\psi_{4,1}-2t'\psi_{3,0}, 
\eea
In these equations, substituting the form of the eigenfunction $\psi_{n,l}$ given in Eq.~\eqref{eq:psi} and demanding $r_{1,1}=0$ gives us the following equations for the locus of points in $(t',\ep_0,E)$-space on which the reflection probability is zero: 
\bea
2\ep_0~=~E, && \ep_0^2+t'^2 \Big(\f{8}{\sqrt{17}+1}\Big) = 1
\eea

\subsection{Larger values of $N$}

\begin{figure}
 \centering
 \includegraphics[width=8cm]{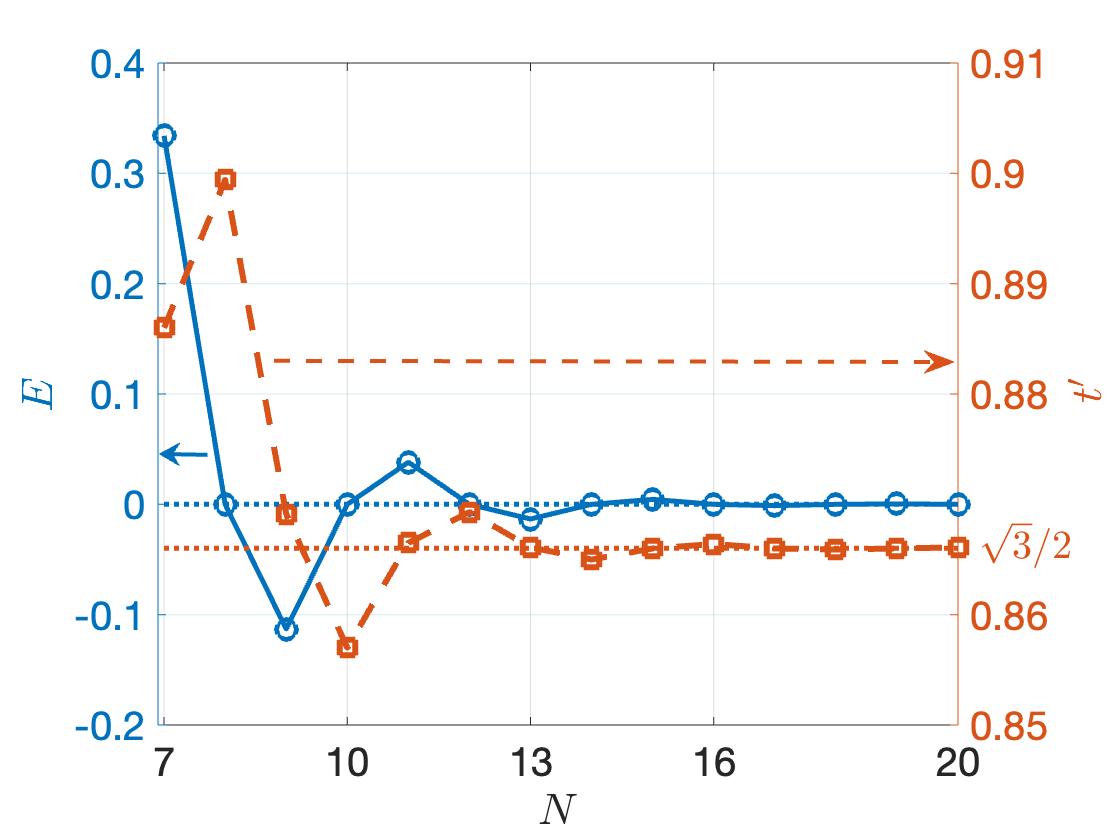}
 \caption{Location of zero-reflection points as a function of $N$ tabulated in Table~\ref{tab:vsN}. It can be seen that in the limit of large $N$, 
 zero-reflection point approaches $(E,t')=(0,\sqrt{3}/2)$. }
 \label{fig:ZRPvsN}
\end{figure}

The locations of the zero-reflection amplitude in the $(k, t')$-space, obtained numerically for $N \ge 7$, are summarized in Table~\ref{tab:vsN} and Fig.~\ref{fig:ZRPvsN}. An inspection of the data reveals a clear even--odd effect in the dependence on $N$.

For even values of $N$, the reflection probability $R$ vanishes at $k = \pi/2$, provided $t'$ is appropriately tuned to a value close to $\sqrt{3}/2$. In contrast, for odd values of $N$, the value of $k$ at which $R = 0$ exhibits oscillations about $\pi/2$, with the amplitude of these oscillations decreasing as $N$ increases for $t'=\sqrt{3}/2$.

A complementary behavior is observed for the hopping parameter $t'$. For odd $N > 10$, the condition $R = 0$ occurs at $t' \approx 0.8660$, whereas for even $N$, the corresponding values of $t'$ oscillate around $0.8660$, again with a diminishing amplitude as $N$ increases.

\begin{table}[htb]
\renewcommand{\arraystretch}{1.2}
\begin{tabular}{ |c|c|c|c| } 
 \hline
$N$& $E$& $t'$ & $R$ \\ 
 \hline
 7& 0.3340 & 0.8860 & $1.3159\times 10^{-9}$ \\ 
 \hline
 8& 0.0000 & 0.8994 & $2.7493\times 10^{-9}$ \\ 
 \hline
 9& -0.113 & 0.8691& $2.1046\times 10^{-9}$ \\ 
 \hline
 10& 0.0000 & 0.8570 & $1.2262\times 10^{-10}$ \\ 
 \hline
 11 & 0.0382 & 0.8665 & $9.1902\times10^{-10}$ \\ 
 \hline
 12 & 0.0000 & 0.8693 & $1.5413\times10^{-15}$ \\ 
 \hline
 13 & -0.0126 & 0.8661 & $3.5777\times10^{-10}$ \\ 
 \hline
 14 & 0.0000 & 0.8650 & $1.6376\times10^{-10}$ \\ 
 \hline
 15 & 0.0045 & 0.8660 & $1.4159\times10^{-9}$ \\ 
 \hline
 16 & 0.0000 & 0.8664 & $1.663\times10^{-10}$ \\ 
 \hline
 17& -0.0012 & 0.8660 & $7.3307\times10^{-10}$ \\ 
 \hline
 18 & 0.0000 & 0.8659 & $2.2536\times10^{-10}$ \\ 
 \hline
 19 & 0.0004 & 0.8660 & $3.1643\times10^{-11}$ \\ 
 \hline
 20 & 0.0000 & 0.8661 & $2.2649\times10^{-11}$ \\ 
 \hline
\end{tabular}
\caption{Location of the zero-reflection amplitude in $(E,t')$-space evaluated numerically for various values of $N$, taking $\ep_0 = 0$ and $t=1$. The last column shows the reflection probability $R$ for the values of $N$, $k$ and $t'$ shown in the other columns. The number of digits on the right-side of decimal point shows the precision.} \label{tab:vsN}
\end{table}

In the limit $N \to \infty$, both even and odd sequences converge to a common point, with vanishing 
reflection occurring at $k = \pi/2$ and $t' = \sqrt{3}/2$. We now study how the reflectionless
point shifts in the $(t',k)$ plane when $N$ is large.

For even values of $N$, it turns out that the
reflectionless point remains at $k=\pi/2$ ($E=0$), but
$t'$ shifts by a small amount which we can calculate
as follows.
First, we note that at $k = \pi/2$, Eq.~\eqref{lpm} implies that
\bea \lm_{\pm} &=& - i \al ~\pm~ i ~\sqrt{1 ~+~ \al^2},
\non \\
{\rm where}~~ \al &=& \f{1}{2t'}. \label{lpm2} \eea
For $t' > 0$, we see that $|\lm_+| < 1 < |\lm_-|$.
For large $N$, the dominant term in
Eq.~\eqref{cond3} leads to
the condition
\beq 2 t' ~\lm_+ ~=~ i, \eeq
whose solution is given by $t' = \sqrt{3}/2$ and $\lm_+ = i/\sqrt{3}$. Next, we compute the correction due to the sub-dominant terms. For $N=2m$, Eq.~\eqref{cond3} implies that
\beq 2 t' \lm_+ ~\left( \f{1 ~+~ \lm_+^{2m-2}}{1 ~+~ \lm_+^{2m}} \right) ~=~ i. \eeq
For large $m$, we can set $\lm_+ = i/\sqrt{3}$ in
the terms $\lm_+^{2m-2}$ and $\lm_+^{2m}$ which are small, but
we keep the exact expression for $2t' \lm_+ =
i (\sqrt{1 + 4t'^2} - 1)$. This gives
\beq \sqrt{4t'^2 + 1} - 1 ~=~ 1 ~-~ \f{4}{3} ~
\left( \f{i}{\sqrt{3}} \right)^{N-2}. \label{cond5} 
\eeq
We note that the factor of $i^{N-2}$ in the correction term 
is real since $N$ is even; however it oscillates in 
sign as $N$ increases in steps of 2.
We therefore see that for large $N$, the deviation of $t'$ from $\sqrt{3}/2$
oscillates in sign and also goes to zero exponentially as
$(1/\sqrt{3})^N$.

For odd values of $N=2m+1$, we find that the
location of the reflectionless point shifts 
slightly from both $t' = \sqrt{3}/2$ and $k = \pi/2$. An analysis similar to the one given above
for even $N$ leads to the condition, 
\bea && \sqrt{4t'^2 - e^{-i2k}} ~-~ i e^{-ik} \nn \\
&& = ~-i e^{ik} ~\left[ 1 ~-~ \f{4}{3} ~
\left( \f{i}{\sqrt{3}} \right)^{N-2} \right]. \label{cond6}
\eea
Note that the factor of $i^{N-2}$ in the correction term 
is imaginary since $N$ is odd; hence the
value of $k$ must necessarily differ from $\pi/2$.
Eqs.~\eqref{cond6} implies that
for large $N$, the deviations of $t'$ and $k$
from $\sqrt{3}/2$ and $\pi/2$ respectively both go to zero exponentially with $N$.

\section{Topological Characterization}
\label{sec6}

The zero-reflection points in the $(E, t')$ parameter space at a fixed $\epsilon_0$ are associated with robust topological features. Specifically, the phase of the reflection amplitude evaluated along a closed loop encircling a zero-reflection point $(E_0, t'_0)$ exhibits a winding of $2\pi$. This behavior is a generic property of complex functions of two real variables and is fundamentally rooted in the Cauchy argument principle of complex analysis~\cite{churchillnbrown}. Analogous topological features appear in  optical systems, where the zero-reflection condition of a Brewster-reflected postparaxial beam produces phase singularities characterized by a quantized phase winding along a closed loop~\cite{nirmal2021}.

In general, if the phase accumulated along a closed contour is $2\pi W_n$, the enclosed region contains at least $|W_n|$ zeros of the complex-valued reflection amplitude. Here, $W_n$ is defined as the winding number for an incident electron from the wire labeled by $n$. Due to the symmetry of the junction, the scattering properties are independent of the choice of lead, rendering all $W_n$ equal to one another. We therefore identify this integer as a universal topological invariant for the junction. To confirm the existence of these zeros, we compute the winding number for representative loops in $(E, t')$-space. A nonzero winding number provides a rigorous signature that the reflection amplitude vanishes at least once within the enclosed region. This topological characterization ensures that while numerical approximations may deviate slightly from the exact coordinates, the true zeros are topologically protected and must reside in the immediate vicinity of the numerically identified points.

We illustrate this for the case of $N=5$ with $\epsilon_0 = 0$. In this regime, we identify two zero-reflection points in the $(E, t')$ plane at $(-1, 1)$ and $(1, -1)$. These points possess winding numbers of $W_n = 1$ and $W_n = -1$, respectively, allowing them to be interpreted as a vortex and an antivortex. As $|\epsilon_0|$ increases toward unity, the vortex and antivortex approach one another in parameter space. Upon reaching $|\epsilon_0| = 1$, they undergo pair annihilation, after which the zero-reflection points vanish [see Fig.~\ref{N5}]. This annihilation occurs at the critical points $(t', \epsilon_0, E) = (0, \pm 1, \pm 2)$. Consequently, the trajectory of zero-reflection points in $(E, t', \epsilon_0)$-space is continuous except at these specific annihilation coordinates.

\section{Effects of On-Site Disorder}
\label{sec7}

\begin{figure}
 \includegraphics[width=7.9cm]{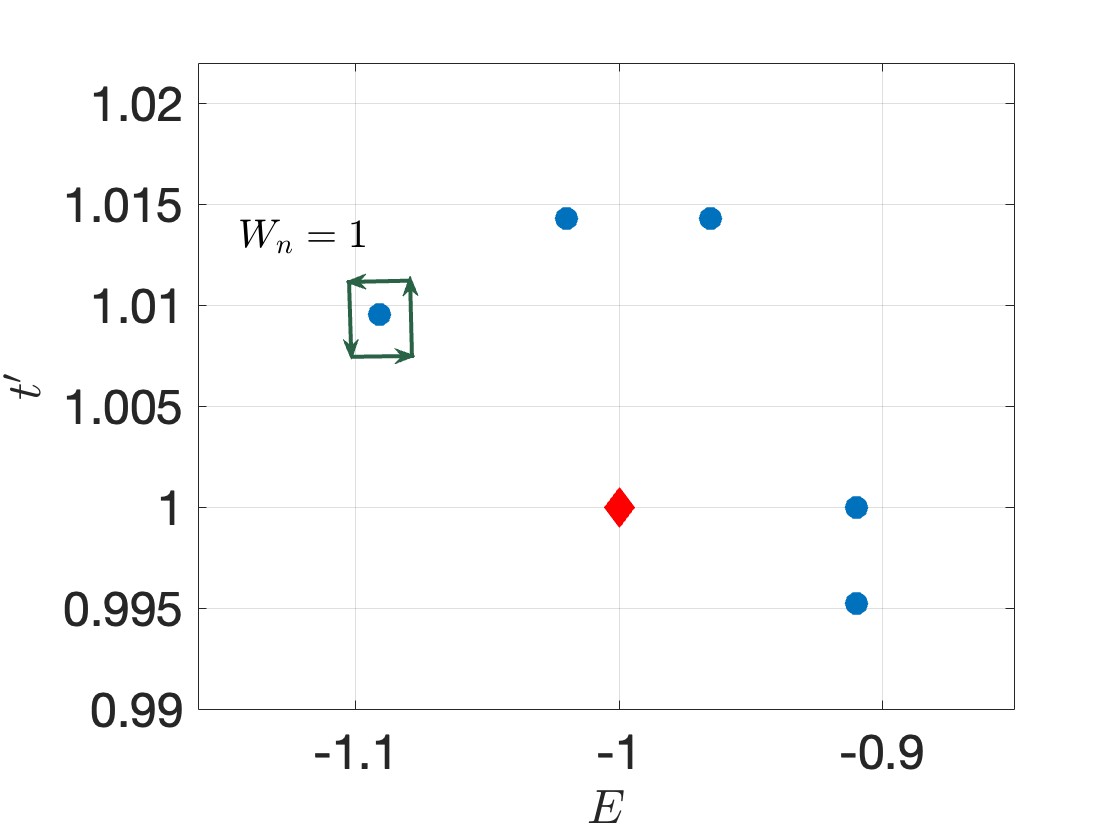}
 \caption{Zero-reflection points for $N=5$. The red diamond denotes the location of the zero-reflection point in the system without disorder ($w=\epsilon_0=0$). Blue circles indicate the shifted, lead-dependent locations of the zero-reflection points for a specific configuration of on-site disorder with strength $w=0.1$. The disorder is drawn from a uniform distribution in the interval $[-w/2, w/2]$. Note that the disorder lifts the degeneracy of the zeros, causing them to occupy distinct positions in the $(E, t')$ plane for different incident wires.}
 \label{fig:dis5}
\end{figure}

We now examine the stability of the zero-reflection points in the presence of on-site disorder at the central junction. On-site disorder is introduced by assigning random energies to the sites of the polygonal junction, drawn from a uniform distribution in the interval $[-w/2, w/2]$, where $w$ denotes the disorder strength. Under weak disorder, the zero-reflection points in the $(E, t')$ parameter space are slightly displaced but do not disappear. Crucially, the disorder breaks the $N$-fold rotational symmetry of the junction, which lifts the degeneracy of the reflection zeros across different leads. Consequently, the zero-reflection points associated with incidence from different wires no longer coincide but instead reside at distinct, lead-dependent locations. This behavior is illustrated in Fig.~\ref{fig:dis5} for $N=5$ with a disorder strength of $w=0.1$ and a specific configuration: $(-0.0464, 0.0349, 0.0434, 0.0179, 0.0258)$.

The robustness of the zero-reflection condition against weak perturbations is a direct consequence of the underlying topological structure of the reflection amplitude. Since the winding number $W_n$ is an integer-valued invariant, the zeros are topologically protected and cannot be eliminated by small perturbations that merely shift their coordinates in parameter space. However, in the strong-disorder regime, the behavior changes qualitatively. As $w$ increases, the zero-reflection points for certain incident channels may eventually be destroyed—for instance, through pair annihilation—while they may persist for others, leading to channel-selective reflectionless transport.

\section{Effects of unequal hopping amplitudes in quantum wires}\label{sec:uneqhop}

\begin{figure}
 \includegraphics[width=7.9cm]{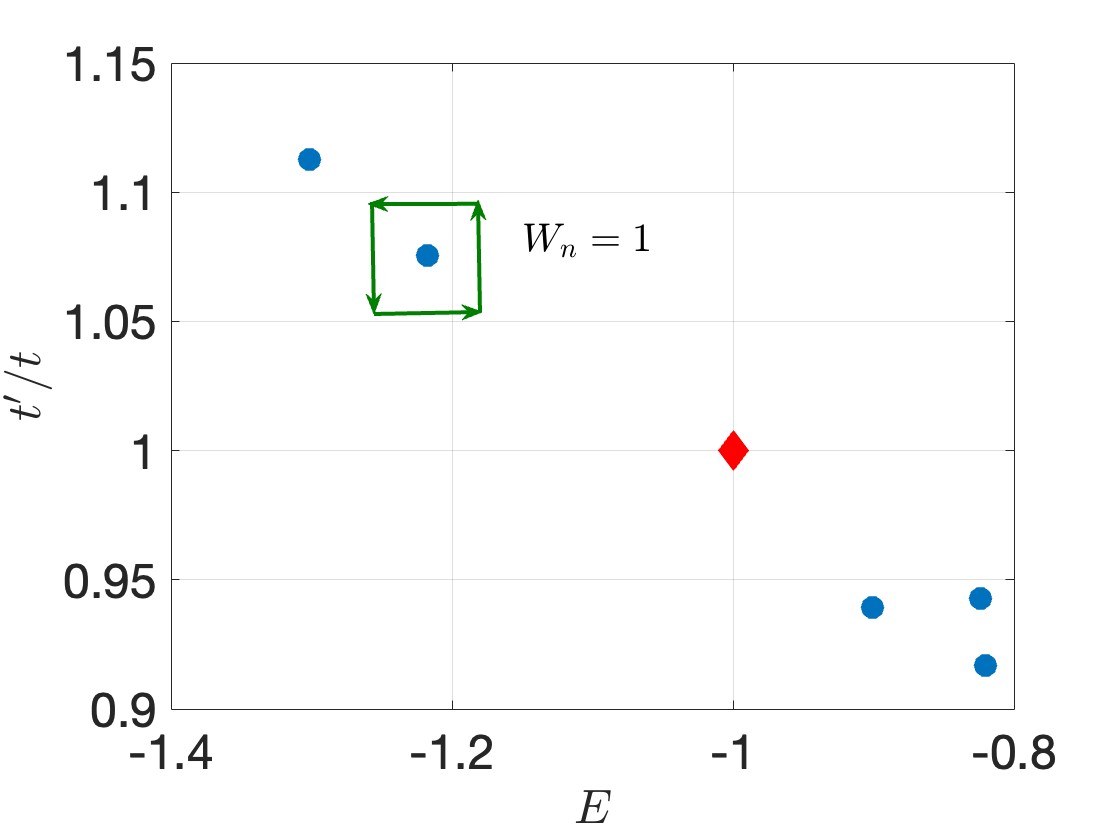}
 \caption{Locations of zero-reflection points in the $(E, t')$ parameter space for a five-wire junction with unequal lead hopping amplitudes $(0.9171, 0.9525, 1.0602, 0.9058, 1.0858)$ at $\epsilon_0 = 0$. Due to the broken five-fold rotational symmetry, waves incident from different wires satisfy the zero-reflection condition at distinct coordinates in $(E, t')$-space. Blue data points mark the zero-reflection positions for the disordered hopping sequence, whereas the red diamond indicates the symmetric reference point where all lead hopping amplitudes equal $1$. Despite these positional shifts, each zero-reflection point remains topologically protected with a winding number of $W_n = -1$.}\label{fig:hoppingdis}
 \end{figure}

In this section, we study the case where the hopping amplitudes differ across the quantum wires while remaining uniform within each individual wire. For a five-wire junction, we select the wire hopping amplitudes randomly from the interval $[1-\delta t, 1+\delta t]$, while fixing the hopping amplitude within the central polygon at $t'$. Setting the on-site energy $\epsilon_0$ to be uniform across all polygon sites, we map the locations of zero-reflection points in the $(E,t')$-space for waves incident from different wires. Although the five-fold rotational symmetry is broken, zero-reflection points persist, albeit at modified locations. Figure~\ref{fig:hoppingdis} shows the locations of zero-reflection points for the representative wire hopping sequence $(0.9171, 0.9525, 1.0602, 0.9058, 1.0858)$ for $\ep_0=0$. Each zero-reflection point remains characterized by a topological winding number of $W_n = -1$.
 
\section{Effects of time-reversal symmetry breaking}
\label{sec8}

In this section, we consider what happens if time-reversal
symmetry is broken at an $N$-wire junction, for example, by
introducing a magnetic flux through the $N$-sided polygon.
For $N=3$ and 4, we find that a flux can indeed give rise
to reflectionless points which are not covered by the
general discussion in Sec.~\ref{sec2} of junctions which 
are symmetric under time-reversal and reflection about one 
of the wires. We discuss these two cases below.

\subsection{$N=3$}

We first consider a triangle with vertices labeled 1, 2
and 3 analogous to the $N=5$ case depicted in Fig.~\ref{fig1}. Setting
$t=1$ and $\ep_0 = 0$ in Eq.~\eqref{eq:ham}, we will study
the condition for having zero reflection for an electron
incident on one of the wires with an energy $E=-2 \cos k$, where $0 < k < \pi$.
To describe the effect of a magnetic flux $\Phi$ 
through the triangle, we follow the Peierls prescription
of introducing appropriate phases in the hopping amplitudes.
Namely, the hopping amplitude 
between nearest neighbors in the anticlockwise direction
(i.e., between pairs of sites $1 \to 2 \to 3 \to 1$) will 
have a phase $\phi/3$, where $\phi = q \Phi/(\hbar c)$ is 
the Aharonov-Bohm phase picked up by the wave function of 
electrons going around the triangle (here $q$ is the
charge of an electron, and $c$ is the speed of light).
The nearest-neighbor hopping amplitude in the clockwise 
direction will then have a phase $- \phi/3$. Since $\phi$
is a periodic variable, we will assume that $- \pi \le \phi
\le \pi$.

For an an electron incident on wire 1 and zero reflection, $r_{1,1} = 0$, Eqs.~\eqref{cond2} take
the form
\bea && t' ~(e^{-i \phi/3} ~t_{2,1} ~+~ e^{i \phi/3} ~t_{3,1}) ~-~ e^{ik} ~=~ 0, \non \\
&& e^{i \phi/3} ~t' ~+~ e^{-i \phi/3} ~t' ~t_{3,1} ~-~ e^{-ik} ~t_{2,1} ~= 0, \non \\
&& e^{-i \phi/3} ~t' ~+~ e^{i \phi/3} ~t' ~t_{2,1} ~-~ e^{-ik} ~t_{3,1} ~= 0. \non \\ 
\label{flux1} \eea
These equations can be written in the matrix form
\bea && M_3 ~\left( \begin{array}{c}
1 \\
t_{2,1} \\
t_{3,1} \end{array} \right) ~=~ \left( \begin{array}{c}
0 \\
0 \\
0 \end{array} \right), ~~~{\rm where} \non \\
&& M_3 ~=~ \left( \begin{array}{ccc}
- e^{ik} & t' e^{-i \phi/3} & t' e^{i \phi/3} \\
t' e^{i \phi/3} & - e^{-i k} & t' e^{-i \phi/3} \\
t' e^{-i \phi/3} & t' e^{i \phi/3} & - e^{-ik}
\end{array} \right). \non \\
\eea
This will have a solution if ${\rm det} (M) = 0$. This gives
the condition
\beq 2 (t')^3 \cos (\phi) ~+~ 2 (t')^2 \cos k ~+~ e^{-ik} 
((t')^2 - 1) ~=~ 0. \eeq
Since $\sin k \ne 0$, the imaginary part of
the above expression being zero implies that we must have
\beq t' ~=~\pm 1. \eeq
The real part being zero then implies that 
\beq \cos k ~\pm \cos \phi = 0 \eeq
for $t' = \pm 1$ respectively.

We now assume that $t' = 1$. Given the ranges of $k$ and 
$\phi$ specified above, we find that $t_{2,1} = 0$
and there is complete transmission from wire 1 to wire 3 
with $t_{3,1} = - e^{i 2 \phi/3}$ if
\beq -\pi < \phi < 0~~ {\rm and}~~ k = \pi + \phi.
\label{fluxcond1} \eeq
By cyclic symmetry, there will also be complete
transmission from wire 3 to wire 2 and from wire 2 to wire 1
when Eq.~\eqref{fluxcond1} is satisfied. The $S$-matrix 
for this system takes the form
\beq
{\cal S}_3 ~=~ - e^{i 2 \phi/3} ~\left( \begin{array}{ccc} 
0 & 1 & 0 \\
0 & 0 & 1 \\
1 & 0 & 0 \end{array} \right).
\label{s3} \eeq

Similarly, we find that if
\beq 0 < \phi < \pi ~~ {\rm and}~~ k = \pi - \phi,
\label{fluxcond2} \eeq
then an electron incident on wire 1 will be
completely transmitted to wire 2, from wire 2 to wire 3,
and from wire 3 to wire 1, and the $S$-matrix is given by 
\beq
{\cal S}'_3 ~=~ - e^{-i 2 \phi/3} \left( \begin{array}{ccc} 
0 & 0 & 1 \\
1 & 0 & 0 \\
0 & 1 & 0 \end{array} \right).
\label{s3p} \eeq

The reflectionless points discussed above have a topological
significance. If we hold $\phi$ fixed, we find that a small
closed curve in the $(k,t')$ space which encircles the
reflectionless point at $k= \pi \pm \phi$ and $t' = 1$ 
has a non-zero winding number equal to $1$.

\subsection{$N=4$}

Next, we consider a square with vertices labeled 1, 2, 3
and 4 analogous to the $N=5$ case depicted in Fig.~\ref{fig1}. 
We put a phase of $- \phi/4$
and $\phi/4$ for nearest-neighbor hopping amplitude in the
clockwise and anticlockwise directions respectively,
so that the Aharonov-Bohm phase picked up in the anticlockwise direction is $\phi$.
We will look for the condition for having zero reflection
for an electron incident on wire 1. Eq.~\eqref{cond2}
then gives the following matrix equation
\begin{widetext}
\bea && M_4 ~\left( \begin{array}{c}
1 \\
t_{2,1} \\
t_{3,1} \\
t_{4,1} \end{array} \right) ~=~ \left( \begin{array}{c}
0 \\
0 \\
0 \\
0 \end{array} \right), ~~{\rm where~~} 
M_4 ~=~ \left( \begin{array}{cccc}
- \ep_0 - e^{ik} & t' e^{-i \phi/4} & 0 & t' e^{i \phi/4} \\
t' e^{i \phi/4} & - \ep_0 - e^{-i k} & t' e^{-i \phi/4} & 0 \\
0 & t' e^{i \phi/4} & - \ep_0 - e^{-ik} & t' e^{-i \phi/4} \\
t' e^{-i \phi/4} & 0 & t' e^{i \phi/4} & - \ep_0 - e^{-i k}
\end{array} \right),
\eea
\end{widetext}
where $k=\cos^{-1}[-E/2]$. This will have a solution if ${\rm det} (M_4) = 0$. 
It can be shown that this holds when 
\bea 
2\ep_0&=&E, \nn \\ 
2t'^2\,\Big|\sin{\f{\phi}{2}}\Big|+\f{E^2}{4}&=&1.
\eea

Setting  $\ep_0=0$ for simplicity, we find that there are
solutions at $k = \pi/2$ and $\phi = \pm \pi/3$.
Choosing $\phi = \pi/3$, we find that $t_{2,1} = - 
e^{-i \pi/4}/ \sqrt{3}$, $t_{3,1} = - 1/\sqrt{3}$ and
$t_{4,1} = 
e^{i \pi/4}/ \sqrt{3}$, which implies that the transmission 
probabilities to wires 2, 3 and 4 are each equal to $1/3$.
Using cyclic symmetry, we can then determine what happens if the
electron is incident on wire 2, 3 or 4. We finally obtain
the $S$-matrix for the system with $t'=1$, $k=\pi/2$ and 
$\phi = \pi/3$ as
\beq {\cal S}_4 = \frac{1}{\sqrt{3}} \left( \begin{array}{cccc}
0 & e^{i \pi/4} & -1 & - e^{- i \pi/4} \\
- e^{-i \pi/4} & 0 & e^{i \pi/4} & - 1 \\
-1 & - e^{-i \pi/4} & 0 & e^{i \pi/4} \\
e^{i \pi/4} & -1 & - e^{- i \pi/4} & 0
\end{array} \right). \\
\label{s4} \eeq

Similarly, for $t'=1$, $k=\pi/2$ and $\phi = - \pi/3$, we find
that the $S$-matrix is given by
\beq {\cal S}'_4 = \frac{1}{\sqrt{3}} \left( \begin{array}{cccc}
0 & - e^{-i \pi/4} & -1 & e^{i \pi/4} \\
e^{i \pi/4} & 0 & - e^{-i \pi/4} & - 1 \\
-1 & e^{i \pi/4} & 0 & - e^{-i \pi/4} \\
- e^{-i \pi/4} & -1 & e^{i \pi/4} & 0
\end{array} \right), \\
\label{s4p} \eeq
which is the transpose of ${\cal S}_4$ in Eq.~\eqref{s4}.

\subsection{$N=4$ and $\phi=\pi$}\label{sec:pi4}

The $\pi$-flux junction with $N=4$ and $\ep_0 = 0$
yields striking scattering properties that can be derived analytically using the open-system retarded Green's function formalism~\cite{datta2005quantum}.
 Although $\phi = \pi$ corresponds to a time-reversal symmetric configuration (since $\phi = \pm \pi$ are identical), we note that this case lies beyond the scope of previous discussions of $N = 4$ junctions with dihedral symmetry in Secs.~\ref{N4-1} and \ref{N4-2}. This is because for any polygonal junction with even $N$, a flux $\phi = \pi$ necessarily breaks the dihedral symmetry down to cyclic symmetry (assuming identical hopping amplitudes on all the bonds). For this reason, we study this case in the present section.
{ The effective Hamiltonian of the four-site junction coupled to identical semi-infinite tight-binding leads is given by $H_{\text{eff}} = H_{4\pi} + \Sigma_L$, where the isolated central ring Hamiltonian with a symmetrically distributed $\pi$-flux  is parameterized as
\begin{equation}
H_{4\pi} = -t' \begin{pmatrix} 0 & e^{-i\pi/4} & 0 & e^{i\pi/4} \\ e^{i\pi/4} & 0 & e^{-i\pi/4} & 0 \\ 0 & e^{i\pi/4} & 0 & e^{-i\pi/4} \\ e^{-i\pi/4} & 0 & e^{i\pi/4} & 0 \end{pmatrix}.
\end{equation}
Because the four leads are structurally identical, the self-energy matrix is proportional to the identity operator, $\Sigma_L = \Sigma_0 I$, where $\Sigma_0 = - e^{ik}$ is the self-energy of an isolated one-dimensional tight-binding lead evaluated at the incident energy $E = -2\cos k$.

The full retarded Green's function of the junction can be expressed in terms of an effective complex energy variable $z = E - \Sigma_0$ as
\begin{equation}
G^R(E) = (E I - H_{4\pi} - \Sigma_0 I)^{-1} = (z I - H_{4\pi})^{-1}.
\end{equation}
Substituting the explicit forms of $E$ and $\Sigma_0$ yields a compact expression for $z = - e^{-ik}$. 

To evaluate the matrix elements analytically, we expand the resolvent using the algebraic identity $(z I - H_{4\pi})^{-1} = (z I + H_{4\pi})(z^2 I - H_{4\pi}^2)^{-1}$. The isolated Hamiltonian has a highly symmetric eigenspectrum with doubly degenerate eigenvalues at $\pm \sqrt{2}t'$. Consequently, its square is strictly proportional to the identity matrix, $H_{4\pi}^2 = 2{t'}^2 I$, an energy-independent constant. Noting that the diagonal elements of $H_{4\pi}$ vanish, the local Green's function at the injection site strictly evaluates to
\begin{equation}
G_{11}^R = \frac{z}{z^2 - 2{t'}^2} = \frac{- e^{-ik}}{ e^{-2ik} - 2{t'}^2}.
\end{equation}

The reflection amplitude $r_k$ is governed by the standard scattering matrix relation $r_k = 1 - i\Gamma G_{11}^R$, where $\Gamma = -2\, \text{Im}(\Sigma_0) = 2 \sin k$ represents the hybridization escape rate into the continuum. Substituting $G_{11}^R$ yields
\begin{equation} 
r_k = 1 + \frac{2i  \sin k \, e^{-ik}}{  e^{-2ik} - 2{t'}^2} =  \frac{1 - 2{t'}^2}{  e^{-2ik} - 2{t'}^2} . \label{eq:rk_pi_flux}
\end{equation}
Equation~\eqref{eq:rk_pi_flux} is an important result. It shows a zero-reflection point with $r_k = 0$ can only exist when the numerator vanishes. This requires $2{t'}^2 = 1$, or equivalently, $t' = 1/\sqrt{2}$. Because the numerator is completely independent of the incident momentum $k$, this specific tuning of the junction hopping guarantees an extraordinary result that a zero-reflection amplitude, $r_k = 0$, occurs for all $k$. This implies that an arbitrary wave packet
that is incident on any wire will not be reflected 
back at all.

The cross-terminal transmission amplitudes follow directly from the off-diagonal elements of $G^R(E)$. Because the $\pi$-flux imposes a destructive interference node at the opposite terminal, $(H_{4\pi})_{31} = 0$, the corresponding Green's function element vanishes, $G_{31}^R = 0$. This  guarantees $t_{3,1} = 0$ independent of the coupling $t'$ or the momentum $k$. 

Conversely, the transmission into the adjacent leads depends explicitly on the junction hopping amplitude $t'$ and the specific flux distribution. Using $(H_{4\pi})_{21} = -t' e^{i\pi/4}$ and $(H_{4\pi})_{41} = -t' e^{-i\pi/4}$, we find
\begin{equation}
G_{21}^R = \frac{-t' e^{i\pi/4}}{  e^{-2ik} - 2{t'}^2}, \quad G_{41}^R = \frac{-t' e^{-i\pi/4}}{  e^{-2ik} - 2{t'}^2}.
\end{equation}
The generic transmission amplitudes into leads 2 and 4 are thus given by
\bea 
t_{2,1} &=& -i\Gamma G_{21}^R = \frac{2i t' e^{i\pi/4} \sin k}{  e^{-2ik} - 2{t'}^2}, \nn \\ 
 t_{4,1} &=& -i\Gamma G_{41}^R = \frac{2i t' e^{-i\pi/4} \sin k}{  e^{-2ik} - 2{t'}^2}.
\eea
In the limit $t' = 1/\sqrt{2}$, the denominator simplifies exactly to $-2i \sin k \, e^{-ik}$, allowing the transmission amplitudes to reduce to an optimal beam-splitting form dressed by the local phase structure of the junction,
\begin{equation}
t_{2,1} = -\frac{1}{\sqrt{2}} e^{i(k + \pi/4)}, \quad t_{4,1} = -\frac{1}{\sqrt{2}} e^{i(k - \pi/4)}.
\end{equation}

Next, we characterize the underlying topological nature of this broadband zero-reflection phenomenon by introducing an alternating on-site potential across the central ring. We assign staggered on-site energies $\ep_s, -\ep_s, \ep_s, -\ep_s$ to the four sites of the polygon. As established above, the coordinate defined by $\ep_s = 0$ and $t' = 1/\sqrt{2}$ acts as a perfect zero-reflection point for any incident momentum $k$. Next, we examine the phase evolution of the reflection amplitude. Along a closed contour in the $(\ep_s, t')$ parameter space that encircles this zero-reflection point $(0,1/\sqrt{2})$, the phase of reflection amplitude $r_{11}$ accumulates a winding of precisely $2\pi$ for any $k$. This corresponds to a non-trivial winding number of $1$, demonstrating that the broadband reflectionless transport in the $\pi$-flux junction is  characterized by the
parameter-space topology. In the space of $(\epsilon_s,t',E)$, the locus of zero-reflection points is a straight line given by $\epsilon_s=0$, $t'=1/\sqrt{2}$.

\section{Zero-Reflection Points for $N=2$}
\label{sec:N2}

Consider a one-dimensional tight-binding chain with a uniform nearest-neighbor hopping amplitude $t=1$. We introduce local impurities by defining the onsite energies at lattice sites $l=0$ and $l=L$ as $\epsilon_0-\delta$ and $\epsilon_0+\delta$, respectively, where $\epsilon_0 \neq 0$. In this configuration, zero-reflection points emerge at isolated locations within the $(E,\delta)$ parameter space. Specifically, these points are strictly confined to the $\delta=0$ axis, occurring at discrete energy values $E$ that are determined by the impurity separation $L$. We note that analogous two-terminal setups exhibiting exact zero-reflection amplitudes have also been demonstrated in photonic lattices~\cite{pal2026}.

The introduction of any finite asymmetry ($\delta \neq 0$) breaks the necessary condition, thereby destroying the zero-reflection points. However, despite their fragility to finite $\delta$, these points possess an underlying topological character in the broader $(E,\delta)$ space. By calculating the phase of the reflection amplitude along a small closed loop enclosing a zero-reflection point, we find that the phase exhibits a quantized winding number of $W = \pm 1$, where the sign is strictly dictated by the sign of $\epsilon_0$. This non-trivial winding number characterizes the zero-reflection points as topological features, even in the simplest two-terminal geometry.

\section{Friedel oscillations in density, and effects of
weak interactions and finite temperatures}
\label{sec9}

In this section, we will discuss an observable signature
of the reflectionless points that have been discussed above.
We will show that Friedel oscillations~\cite{friedel1958,tutto1985,egger1995} in the density
in any wire vanishes when the Fermi momentum $k_F$
coincides with a reflectionless momentum in that wire.
We will then use this result to study what happens when
the electrons interact weakly with each other in the wires~\cite{lal2002,yue1966}. Finally, we discuss the effects of finite temperature. 

\subsection{Friedel oscillations in density}
\label{sec9a}

\begin{figure*}[htb]
\includegraphics[width=5.3cm]{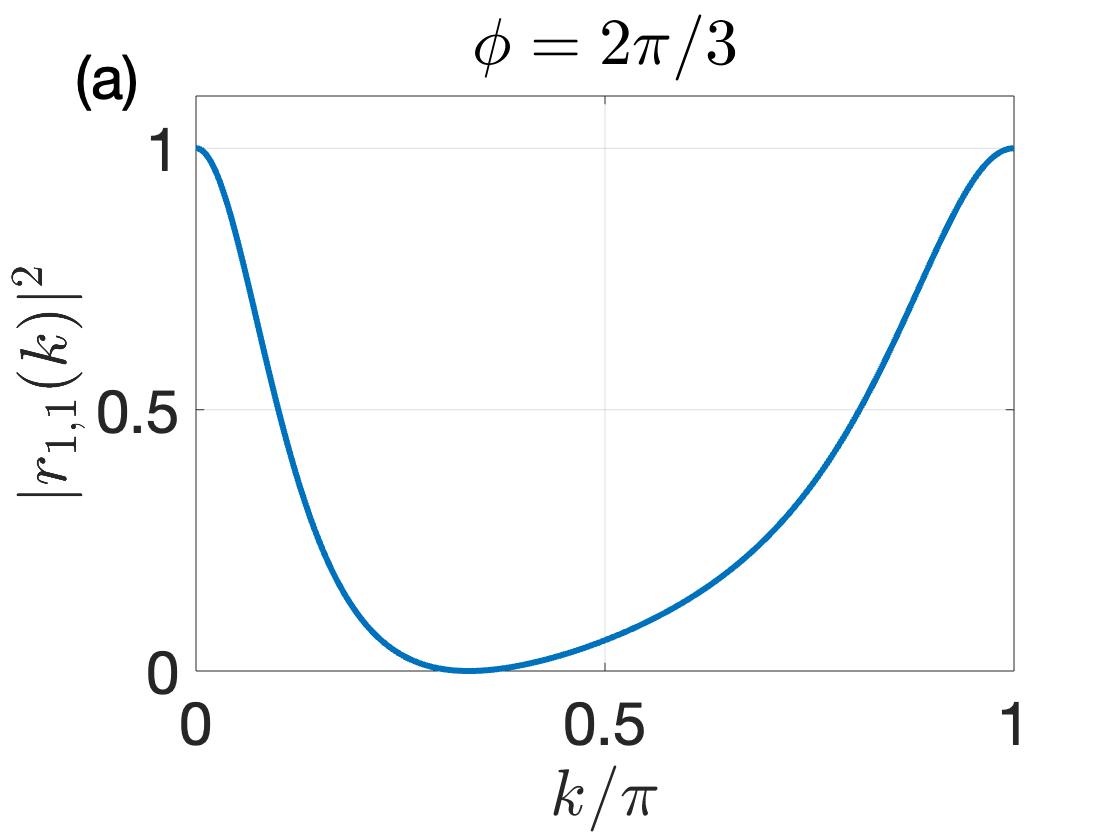}
\includegraphics[width=5.3cm]{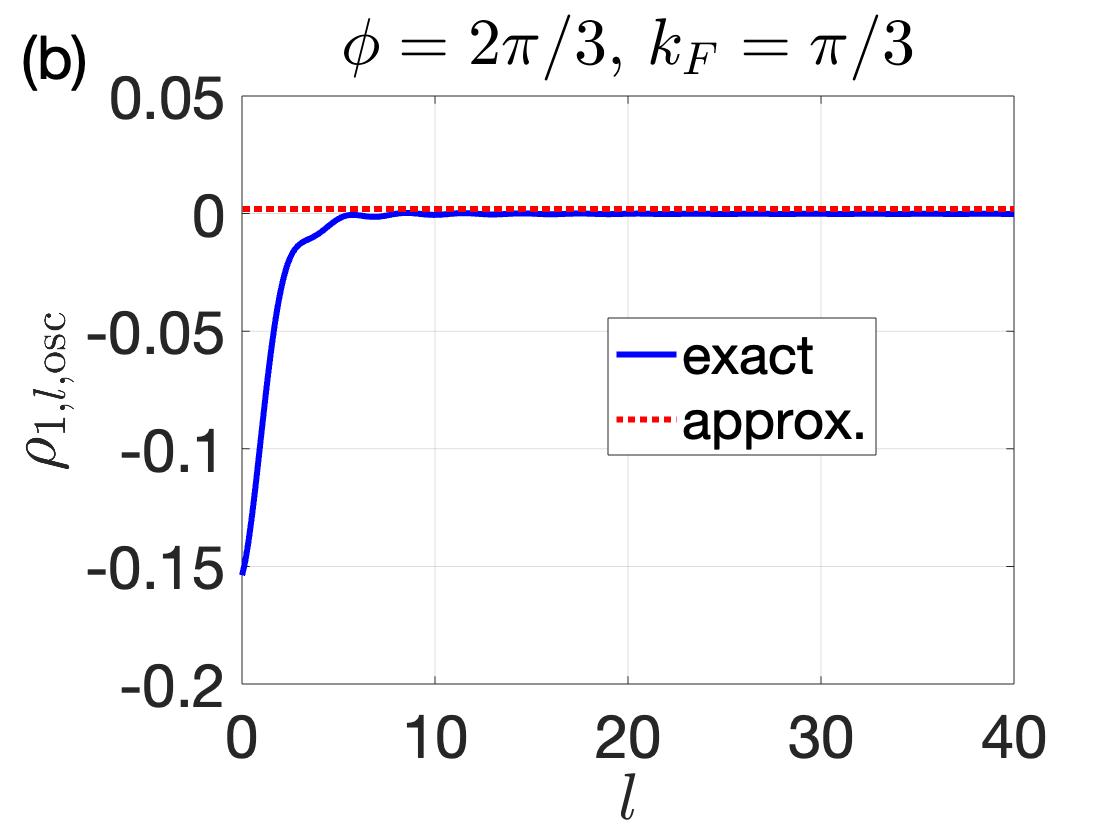}
\includegraphics[width=5.3cm]{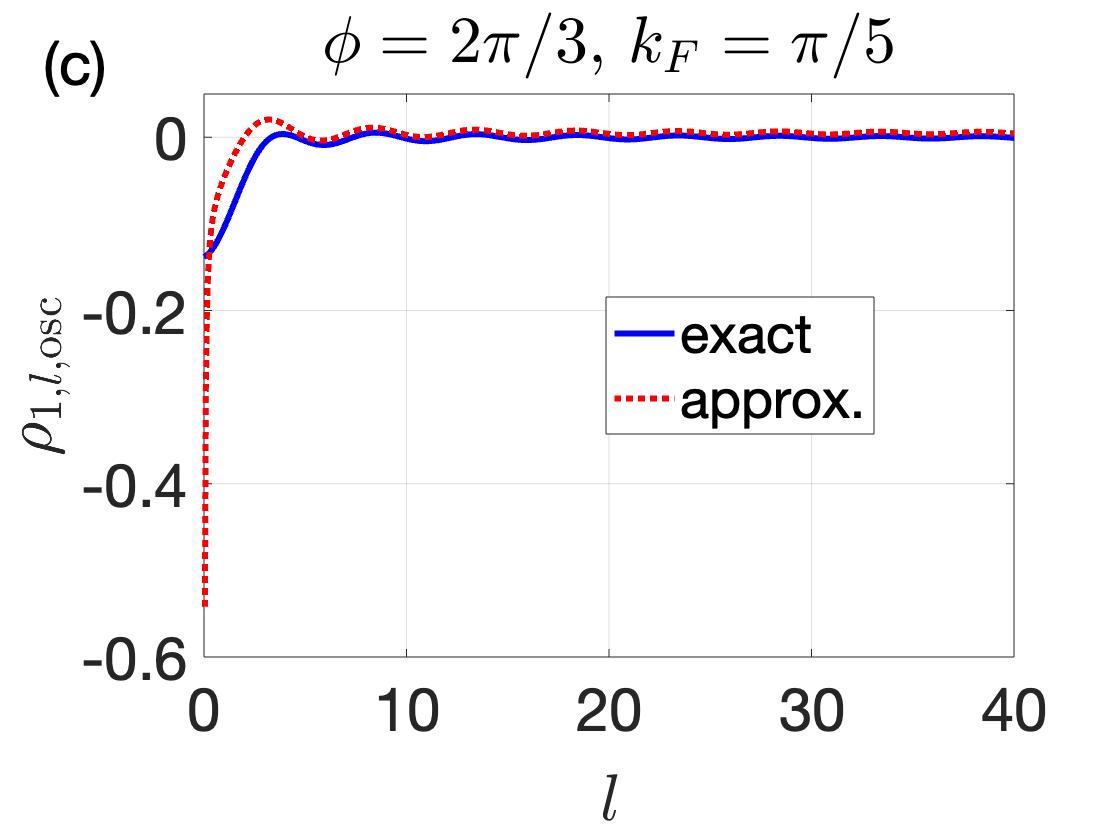}
\caption{Plots for a three-wire junction 
enclosing a flux $\phi = 2\pi/3$. There is no reflection
when $k=\pi/3$. (a) Reflection probability in any wire versus $k$ for the full range $0 < k < \pi$. (b) Oscillatory part of the density in any wire 
versus the site index $l$ for $k_F = \pi/3$ which is a reflectionless point. 
The blue line denotes the exact numerical result,
$\rho_{1,l,osc}$, obtained by integrating over $k$ from $0$ to $k_F$. The red line shows the approximate result, $\rho'_{1,l,osc}$, calculated using only the reflection amplitude at $k=k_F$.
(c) Same plot as in (b) except that $k_F = \pi/5$ where
$r(k_F) \ne 0$. The blue and red lines match very well 
for large values of $l$. The oscillations have a period 
equal to $\pi/k_F = 5$. In plots (b) and (c) the red curves are displaced upwards by $0.002$ and $0.005$ respectively along the $y$-axis for contrast.} 
\label{fig:friedel} \end{figure*}

We begin with a general discussion of Friedel oscillations. Consider an $N$-wire
junction and let the Fermi energy for the system be given
by $E_F = -2  \cos k_F$, so that $k_F$ is the Fermi momentum (this lies in the range $0 < k_F < \pi$ as usual).
At zero momentum, the incoming states in all the wires will
have all possible values of momentum lying in the range
$0 \le k \le k_F$. The density at site $l$ of the $n$-th wire is then given by the integral
\bea && \rho_{n,l} ~=~ \int_0^{k_F} \frac{dk}{2\pi} ~
\sum_{m=1}^N ~|\psi_{n,m,l} (k)|^2, ~~{\rm where} \non \\
&& \psi_{n,m,l} (k) ~\equiv~ t_{n,m} (k) e^{i k l} ~~
{\rm if}~~ m \ne n, \non \\
&& \psi_{n,m,l} (k) ~\equiv~ e^{- i k l} + r_{n,n} (k) 
e^{i k l} ~~ {\rm if}~~ m = n. \non \\
\label{rhonl} \eea
Physically, $\psi_{n,m,l} (k)$ denotes
the contribution to the wave function
at site $l$ on wire $n$ for a wave
incident with unit amplitude with momentum $k$ on wire $m$.
We now concentrate on the first wire, labeled $n=1$.
The unitarity of the scattering matrix ${\cal S} (k)$
implies that 
\beq |r_{1,1} (k)|^2 ~+~ \sum_{m=2}^N |t_{1,m} (k)|^2 ~=~ 1
\eeq
for any value of $k$. Eq.~\eqref{rhonl} then leads to 
\beq \rho_{1,l} ~=~ \frac{k_F}{\pi} ~+~ \int_0^{k_F} 
\frac{dk}{2\pi} ~[r_{1,1} (k) e^{i 2 k l} + r_{1,1}^* (k)
e^{-i 2 k l}]. \label{friedel1} \eeq
We now focus on the second term in Eq.~\eqref{friedel1},
\bea && \rho_{1,l,osc} ~=~ \rho_{1,l} ~-~ \frac{k_F}{\pi} 
\non \\
&& =~ \int_0^{k_F} 
\frac{dk}{2\pi} ~[r_{1,1} (k) e^{i 2 k l} + r_{1,1}^* (k)
e^{-i 2 k l}]. \label{friedel2} \eea
We can use an approximation to analytically evaluate the 
expression in Eq.~\eqref{friedel2}. We assume that 
$r_{1,1} (k)$ is independent of $k$
and is given by its value at $k=k_F$,
which we denote as $r(k_F)$ to simplify the notation. 
Then the integration in Eq.~\eqref{friedel1} approximately gives the oscillatory part of the density as
\beq \rho'_{1,l,osc} ~=~ - \frac{i}{4 \pi l} ~
[r (k_F) e^{i2 k_F l} ~-~ r^* (k_F) e^{-i2 k_F l}], \label{friedel3} \eeq
where we have ignored the contribution from the lower 
limit, $k=0$. We see from Eq.~\eqref{friedel3} that if
$r(k_F) \ne 0$, 
$\rho'_{osc,1,l}$ oscillates with $l$ with a wavelength
equal to $\pi/k_F$, and its magnitude decays as $1/l$ for
large $l$.
If $r(k_F) = 0$, we have to return to Eq.~\eqref{friedel2}
and redo the calculation. Typically, the derivative 
$dr_{1,1} (k)/dk \ne 0$ at $k=k_F$, and we then find that
$\rho_{1,l,osc}$ again oscillates with a wavelength $\pi/
k_F$, but its magnitude now decays as $1/l^2$ for large $l$.

As an example, we consider a three-wire junction with
a flux $\phi$ as discussed in Sec.~\ref{sec8}~A. We choose
$\phi = 2 \pi/3$, which gives a reflectionless point at
$k=\pi/3$ according to Eq.~\eqref{fluxcond2}. This
is shown in Fig.~\ref{fig:friedel} (a). 
Since the system has cyclic symmetry, the
density $\rho_{n,l}$ will have the same form on all the
wires. Fig.~\ref{fig:friedel} (b) shows the numerically
exact (Eq.~\eqref{friedel2}) and approximate (Eq.~\eqref{friedel3}) forms in blue and red respectively for
$k_F = \pi/3$; the approximate form is exactly equal to
zero since $r(k_F) = 0$. Fig.~\ref{fig:friedel} (c) shows 
the numerically
exact and approximate forms in blue and red respectively for
$k_F = \pi/5$. We see that they match extremely well 
beyond about $l=8$. As mentioned above, the oscillations
have a period equal to $\pi/k_F = 5$.

In conclusion, the real-space particle density exhibits prominent Friedel oscillations whose amplitude decays as $1/l$ (where $l$ is the distance from the junction) when the Fermi momentum deviates from the zero-reflection condition. Conversely, at the zero-reflection points, these spatial oscillations decay much more rapidly. Consequently, a systematic measurement of Friedel oscillations as a function of the Fermi momentum $k_F$ provides a viable and non-invasive experimental signature for detecting and mapping these reflectionless points in multi-terminal quantum wire junctions.

\subsection{Effects of weak interactions}
\label{sec9b}

We will now discuss the effects of weak interactions
between the electrons in the wires following the 
method presented in Ref.~\cite{lal2002}. The analysis
is based on a continuum description of the wires; this
is valid when the length scales of interest are all much
larger than the lattice spacing $a=1$. Further, we will
work in the vicinity of the Fermi momentum $k_F$,
where the Fermi velocity of the electrons in the lattice
model is given by $v_F = 2  \sin (k_F)$.
We begin
by introducing a dimensionless interaction parameter
$\al_n$ in wire $n$. In the Tomonaga-Luttinger liquid 
description of spinless electrons with short-range 
density-density interactions in one dimension, the 
Luttinger parameters in wire $n$ are related to $\al_n$ as 
\bea v_n &=& v_F (1 ~-~ \al_n^2)^{1/2}, \non \\
K_n &=& \left( \frac{1 ~-~ \al_n}{1 ~+~ \al_n} \right)^{1/2}, \eea
where $v_n$ is the velocity of the bosonic (particle-hole) excitations, and $K_n$ is the 
Luttinger parameter. For noninteracting electrons, we have
$\al_n = 0$ and $K_n = 1$. For repulsive (attractive) 
interactions, $\al_n > 0 ~(< 0)$ and therefore $K_n < 1 ~(> 1)$ respectively.

We will simplify our notation by denoting the reflection 
amplitude in wire $n$ as $r_{nn} = r_{n,n} (k_F)$ and the 
transmission amplitude from wire $m$ to wire $n$ as $t_{nm} 
= t_{n,m} (k_F)$. These define a scattering matrix $S$ as
usual. Next, we define a diagonal matrix $F$ whose 
entries are given by
\beq F_{nn} ~=~ - ~\frac{1}{2} ~\al_n ~r_{nn}. 
\label{fmat} \eeq
Ref.~\cite{lal2002} then shows that if $| \al_n | \ll 1$
(i.e., for weak interactions), the scattering matrix
${\cal S}$ (and therefore the matrix $F$ whose elements are related
to the diagonal elements of ${\cal S}$) becomes a function of a 
distance scale $L$, and the variation of ${\cal S}$ with $L$
is given by a renormalization group (RG) equation
\beq \frac{d{\cal S}}{dl} ~=~ {\cal S} F^\dagger {\cal S} ~-~ F, \label{rg} \eeq
where $l = \ln (L/a)$. Eq.~\eqref{rg} is only valid to
first order in the $\al_n$'s; these parameters do not
themselves flow under the RG. One can check that 
Eq.~\eqref{rg}  maintains the unitarity condition, ${\cal S}^\dagger S = I$.
We note that the derivation of Eq.~\eqref{rg} crucially
uses the fact that Friedel oscillations in the density (in 
particular, the oscillations of the form $e^{\pm i 2 k_F x}$),
lead to backscattering of electrons between incoming and
outgoing states which lie at the momenta $- k_F$ and
$+k_F$~\cite{yue1966,lal2002}.

Eqs.~\eqref{fmat} implies trivially that the matrix $F$
vanishes if $r_{nn} =
0$ in all the wires. It then follows from Eq.~\eqref{rg} that ${\cal S}$ will not evolve under RG, at least to first order
in $\al_n$. Reflectionless points therefore have the 
remarkable property that they are fixed points of the RG 
flow when the interactions are weak.

Next, we study the stability of the fixed point at $r_{nn}
=0$. As a specific example, we consider the three-wire 
junction with a flux discussed in Sec.~\ref{sec8}~A. We
choose a value of $k_F$ 
which is slightly away from a reflectionless point, so that
$|r_{nn}|$ is non-zero but small. We then numerically study
how ${\cal S}$ evolves under Eq.~\eqref{rg}. Taking all the 
$\al_n$'s to be equal and positive (i.e., repulsive 
interactions), we find that $|r_{nn}|$ increases and
eventually approaches 1 as the RG scale $l$ grows. Hence,
in this system, the reflectionless point is an
unstable fixed point if the interactions are repulsive.
In contrast, if we take the $\al_n$'s to be equal and 
negative (i.e., attractive interactions), we find that 
$|r_{nn}|$ decreases and eventually approaches zero as 
$l$ grows. Thus the reflectionless point is a
stable fixed point for attractive interactions.

\subsection{Effects of finite temperatures}
At finite temperatures, thermal smearing governed by the Fermi-Dirac distribution introduces an energy averaging window of order $k_B T$ around the Fermi energy. Because the zero-reflection points are topologically protected, the reflection amplitude increases continuously and smoothly away from the exact zero-reflection energy. Thus, the junction remains highly transparent as long as $k_B T$ is narrower than the energy width of the zero-reflection valley. Most notably, the broadband zero-reflection condition discovered in the four-terminal junction threaded by a $\pi$-flux occurs for all incident energies. This renders the perfect transmission strictly immune to thermal smearing. Furthermore, while finite temperatures will  damp real-space Friedel oscillations beyond the thermal length scale ($L_T \sim \hbar v_F / k_B T$, where $v_F$ is the Fermi velocity), the characteristic signature of the zero-reflection fixed points -- the rapid spatial decay of these oscillations -- will remain observable within the thermal coherence length.

\section{Summary and Conclusions}
\label{sec10}

In this work, we have systematically investigated ballistic transport through $N$-terminal quantum wire junctions, revealing the existence of exact zero-reflection points. While fully symmetric junctions strictly forbid reflectionless scattering for $N\ge 3$, relaxing the symmetry constraints allows the reflection amplitude to vanish at specific, isolated coordinates within the $(E, t')$-space. Our numerical and analytical evaluations demonstrate a clear even--odd effect in the location of these points, which ultimately converge to $k=\pi/2$ and $t'=\sqrt{3}/2$ in the thermodynamic large-$N$ limit. 

Crucially, we established that these reflectionless coordinates possess a robust topological character, defined by an integer-valued winding number. This topological protection ensures that weak on-site disorder merely shifts the zero-reflection coordinates in parameter space—lifting the lead-dependent degeneracy due to the breaking of $N$-fold rotational symmetry—but does not immediately eliminate them. Annihilation of these points requires stronger perturbations that force the topological 
zero-reflection points (vortex and antivortex pair) to meet and mutually destroy one another. 

Furthermore, we explored the effects of breaking time-reversal symmetry. By introducing a magnetic flux through the junction, we demonstrated that non-trivial reflectionless transport can be engineered even in geometrically constrained $N=3$ and $N=4$ systems. Most notably, in the specific case of an $N=4$ junction threaded by a $\pi$-flux, we identified a broadband zero-reflection condition that yields perfect, energy-independent transmission, thereby significantly extending the versatility and reach of our proposed junction.

From an experimental and interacting perspective, these zero-reflection points leave distinct physical signatures. Specifically, the real-space Friedel oscillations in the particle density decay significantly faster when the Fermi momentum aligns with a zero-reflection point, providing a viable and non-invasive diagnostic tool for detecting these modes in solid-state devices. Additionally, we proved that these reflectionless coordinates serve as fixed points under renormalization group flow when subjected to weak inter-particle interactions. 

Taken together, our results provide a comprehensive theoretical framework for realizing protected, reflectionless transport in noninteracting multi-terminal quantum junctions. These findings open new avenues for designing highly efficient quantum wire networks and manipulating topological scattering phases in scalable nanoelectronic architectures.

\begin{acknowledgments}
We thank Arijit Saha and Nirmal K Viswanathan for stimulating discussions. The authors acknowledge the use of Google Gemini~3.1~Pro for language polishing and structural refinement of the analytical arguments in Section\ref{sec:pi4}. All content was reviewed and verified by all the authors. AS acknowledges the financial support received from the Anusandhan National Research Foundation, India (erstwhile Science and Engineering Research Board, India) under the Core Research Grant (No. CRG/2022/004311), and from the University of Hyderabad. UK acknowledges support from the Physical Research Laboratory, Ahmedabad and the Department of Space, Government of India. 
\end{acknowledgments}

\bibliography{ref_nwirejn}

\end{document}